\documentclass[onecolumn]{article}

\usepackage{graphicx} 
\usepackage{color}

\usepackage{amsmath}
\usepackage{mathtools,bm}
\usepackage[round]{natbib}
\usepackage{xspace}
\usepackage{pifont}   
\usepackage{slashbox,booktabs}

\usepackage{setspace}

\usepackage{times}

\usepackage[T1]{fontenc}

\newcommand{\bfnabla}{\mbox{\boldmath $\bf \nabla$}}

\def\rmd{{\mathrm{d}}}

\newcommand{\tdot}{\,.\hspace{-0.98 mm}\raise.6ex\hbox{.}
                   \hspace{-0.98 mm}\raise1.2ex\hbox{.}\,}

\def\bfc{{\mathbf{c}}}

\def\bff{{\mathbf{f}}}

\def\bfm{{\mathbf{m}}}
\def\bfn{{\mathbf{n}}}

\def\bfr{{\mathbf{r}}}
\def\bfs{{\mathbf{s}}}

\def\bfv{{\mathbf{v}}}
\def\bfw{{\mathbf{w}}}
\def\bfx{{\mathbf{x}}}

\def\bfM{{\mathbf{M}}}

\def\bfT{{\mathbf{T}}}

\begin{document}

\title
{Detectable seismic consequences of the interaction of a primordial black hole with Earth}

\author{
Yang Luo$^{1}$\,, Shravan Hanasoge$^{1,2}$\,, Jeroen Tromp$^{1,3 \star}$ and Frans Pretorius$^{4}$
}
\date{}
\maketitle
\noindent
${}^{1}$ Department of Geosciences, Princeton University, Princeton, NJ~08544, USA \\
${}^{2}$ Max-Planck-Institut f\"{u}r Sonnensystemforschung, Katlenburg-Lindau, Germany\\
${}^{3}$ Program for Applied \& Computational Mathematics, Princeton University, Princeton, NJ~08544, USA \\
${}^{4}$ Department of Physics, Princeton University, Princeton, NJ~08544, USA \\ 
${}^{\star}$ e-mail: jtromp@princeton.edu \\

\begin{abstract}
Galaxies observed today are likely to have evolved from density perturbations in the early universe.
Perturbations that exceeded some critical threshold are conjectured to have
undergone gravitational collapse to form primordial black holes (PBHs) at a range of masses \citep{1967SvA....10..602Z,Hawking1971,CarrHawking1974}.
Such PBHs serve as candidates for cold dark matter \citep{Carretal2010} and their detection would shed light on conditions in the early universe.
Here we propose a mechanism to search for transits of PBHs through/nearby Earth by studying the associated seismic waves.
Using a spectral-element method \citep{Koma99}, we simulate and visualize this seismic wave field in Earth's interior.
We predict the emergence of two unique signatures, namely, a wave that would arrive almost simultaneously everywhere on Earth's free surface and the
excitation of unusual spheroidal modes with a characteristic frequency-spacing in free oscillation spectra.
These qualitative characteristics are unaffected by the speed or proximity of the PBH trajectory.
The seismic energy deposited by a proximal ${M^\mathrm{PBH} = 10^{15}}$~g PBH is comparable to a magnitude ${M_\mathrm{w}=4}$ earthquake.
The non-seismic collateral damage due to the actual impact of such small PBHs with Earth would be negligible. Unfortunately, the expected collision rate is very
low even if PBHs constituted all of dark matter, at ${\sim 10^{-7}~\mathrm{yr}^{-1}}$, and since the rate scales as ${1/M^\mathrm{PBH}}$, fortunately encounters with larger, 
Earth-threatening PBHs are exceedingly unlikely.
However, the rate at which non-colliding close encounters of PBHs could be
detected by seismic activity alone is roughly two orders of magnitude larger
--- that is once every hundred thousand years ---
than the direct collision rate.
\end{abstract}

\section{Introduction}

The determination of the constituents of dark matter is an outstanding problem that continues to receive great attention \citep{Feng:2010gw}.
Constraints derived from the gravitational microlensing survey Exp\'{e}rience de Recherche d'Objets Sombres \citep{Alcocketal1998,Tisserandetal2007}
limit at most 8\% of our galaxy's dark matter halo to be comprised of objects with mass $M^\mathrm{PBH}$ between
$10^{26}$~g and  $2.4 \times 10^{34}$~g (for comparison, Earth's mass is approximately $6\times 10^{27}$~g).
PBHs, which could constitute a significant fraction of these massive objects, are subject to Hawking radiation losses \citep{Hawking1974},
a process by which they emit photons of a characteristic frequency $\propto 1/M^\mathrm{PBH}$.
PBHs with masses $M^\mathrm{PBH} < 10^{15}$~g would thus evaporate within the age of the universe,
and those with $M^\mathrm{PBH} < 10^{17}$~g would emit potentially detectable
Gamma rays \citep{PageHawking1976,Sreekumaretal1998,Carretal2010}.
Based on these criteria, PBHs in the mass range $10^{17}~\mathrm{g} < M^\mathrm{PBH} < 10^{26}~\mathrm{g} $
are unconstrained and the entirety of dark matter in our galaxy could in principle be constituted by them.
One way to narrow these constraints is to search for special oscillations of the Sun (and potentially other stars)
induced by the gravitational interaction with proximal and impacting PBHs.
The impacts of PBHs of the size of asteroids ($M^\mathrm{PBH} = 10^{21}$~g) with the Sun would be detectable
by existing solar observatories, but the rate of occurrence of such events is estimated at a disappointingly low $10^{-8}$ yr$^{-1}$ \citep{KesHan2011}.

The prospect of black holes striking Earth excites considerable public interest.
The PBHs we consider here are of masses small enough that they do not pose an existential threat to our planet, and because they are at the lower end of the allowable mass range, will cause negligible non-seismic effects.
However, for masses the size of asteroids and larger, devastation similar in some manner to the impact of an
asteroid may ensue.
Consider, for example, the Tunguska impact and explosion in Siberia on June 30, 1908.
No comet ash has ever been recovered around this region, leading to the speculation that
it might have occurred due to an impact with a PBH \citep{JackRyan1973}.
Subsequently, this theory was dismissed because of (among other reasons) the lack of secondary signatures around
the exit location of the supposed PBH \citep{BeasTins1974,Burnsetal1976}.
Another example is the putative identification of supersonic waves in recorded seismograms,
hypothesized to have been caused by the impact of a `quark nugget' with Earth \citep{Davidetal2003} on Nov. 24, 1993.
This claim was later falsified when it was shown that the signals were actually due to a hitherto unreported earthquake \citep{Selbyetal2004}.
Quark nuggets (QNs), also termed strangelets, are a hypothetical phase of matter
comprised of roughly equal parts of up, down and strange quarks that have lower binding energy
than iron, and thus may be the ``ultimate'' stable state of matter \citep{witten84}. Because
a QN is comprised of materials at nuclear density and can be of any
mass up to a neutron star mass, it will have a very similar gravitational effect
as a PBH when passing through Earth, and present a similar seismic signature.
However, smaller QNs, of
order $M = 10^{15}$~g, could still produce observable explosions in the atmosphere,
in contrast to PBHs of the same mass. This could be used to distinguish these
two classes of compact object collisions with Earth.

The speed of PBHs in unbound orbits \citep{lisanti_spergel} is on the order of 200--400~km~s$^{-1}$,
indicating a maximum transit time through Earth of less than a minute.
During its passage, a PBH would exert gravitational forces on the whole globe and generate seismic waves
that propagate through Earth's interior.
Due to the fast movement of the PBH compared to typical seismic wave speeds, such an event may be treated effectively as a
supersonic source that generates Mach waves.
The signatures of Mach waves are different from those of earthquake-induced seismic waves, thereby providing a means of detecting PBHs.
Amplitudes of such waves have also been estimated \citep{Khrietal2008}, albeit in a simplistic scenario where Earth is treated as a homogeneous fluid ball.

Gravitational forces in the near-field of the PBH are very strong, leading to non-linear effects such as the cracking of
rocks and heating. However, in this paper we focus on the far-field, treating the PBH as a moving Newtonian
gravitational source to the seismic wave equation \citep{KesHan2011},
which we solve using the publicly available {\texttt{SPECFEM3D\_GLOBE}} software package.
An outline of the rest of the paper is as follows. 
In Section~\ref{section:ge_and_ni} we introduce
the governing equations and describe the numerical solution method. In Sections~\ref{section:results}
and~\ref{section:modes} we discuss the results, in the latter section focusing on the
unique spherical normal modes induced by a PBH. In Section~\ref{section:estimates} we provide
some order-of-magnitude estimates of event rates and non-seismic effects. In Section~\ref{section:recipe}
we suggest a strategy for detection, before concluding in Section~\ref{section:conclusions}.
We relegate to the appendix convergence tests and a few additional simulation results.

\section{Governing Equations and Numerical Implementation}\label{section:ge_and_ni}

\subsection{Governing Equations}
\label{section:governing_equations}
In seismology, the displacement field $\bfs$ of particles within Earth's volume $\Omega$ is governed 
by the seismic wave equation
\begin{equation}
\rho \, \partial^2_t \bfs - \bfnabla \cdot \bfT = \bff \quad,
\label{eq:governing_equation}
\end{equation}
where $\rho$ is the mass density, $\bfT$ the symmetric second-order stress tensor and $\bff$ the source. 
Equation (\ref{eq:governing_equation}) is a linearized version of conservation of linear momentum, which proves to 
be sufficiently accurate for seismic wave propagation in Earth's interior.

In an elastic material, the stress tensor $\bfT$ is determined in terms of the strain via Hooke's law,
\begin{equation}
\bfT = \bfc~{\bm:}~{\bm\varepsilon}  \quad,
\label{eq:constitutive_law}
\end{equation}
where $\bfc$ is the fourth-order elastic tensor with at most 21 independent components, describing the elastic properties of 
the material, and ${\bm\varepsilon}$ the strain tensor
\begin{equation}
{\bm\varepsilon=\frac{1}{2}\left[\bfnabla\bfs+\left(\bfnabla\bfs\right)^T\right]} \quad.
\end{equation}
In an isotropic elastic medium, the fourth-order tensor reduces to 2 independent parameters --- the bulk modulus $\kappa$ and the shear modulus $\mu$. 
Furthermore, the shear modulus $\mu$ vanishes in an acoustic medium (e.g., the outer core).
Therefore, only the parameter $\kappa$ is relevant in fluids.
It is important to note that because of this difference between solids (inner core, crust \& mantle) and
liquids (outer core), the governing equation takes slightly different forms,
as will be further discussed later.
Complications associated with attenuation are readily captured by an absorption band model \citep{DT},
and are accommodated in numerical simulations based on the introduction of memory variables~\citep[e.g.,][]{Koma99}.

In global/regional seismology, the source term $\bff$ represents earthquakes, whereas in exploration seismology, 
active sources, such as man-made explosions, are usually employed.
To study the effects of PBHs, neither source type is relevant.
Following \cite{KesHan2011}, we consider a classical black hole,
for which the source term $\bff$ may be written as the gradient of the gravitational potential $\phi$ of the black hole:
\begin{equation}
\bff = \mbox{} -\rho \, \bfnabla \phi \quad.
\label{eq:driving_force}
\end{equation}
The potential $\phi$ obeys Newton's law of gravitation
\begin{equation}
\label{eq:gravitational_potential}
\phi(\bfx,t)=\mbox{}-\frac{G\,M^\mathrm{PBH}}{~||\bfx-\bfx^\mathrm{PBH}(t)||} \quad,
\end{equation}
where $G$ is the universal gravitational constant and $M^\mathrm{PBH}$ the mass of the PBH.
Note that the gravitational potential (hence the gravitational force) changes with time
as the PBH is fast moving with a trajectory
\begin{equation}
\bfx^\mathrm{PBH}(t)=\bfx^\mathrm{PBH}_0-\bfv^\mathrm{PBH} \, t \quad,
\label{eq:trajectory}
\end{equation}
where $\bfx^\mathrm{PBH}_0$ is its initial position (when we start the simulation) and $\bfv^\mathrm{PBH}$ the velocity of the PBH.
The energy loss of the PBH during its passage through Earth is so small
that the velocity of the PBH remains constant.
Consequently, the source term $\bff$ takes the form
\begin{equation}
\bff(\bfx,t) = \mbox{}-~G\,M^\mathrm{PBH}\,\rho(\bfx)\,\frac{\bfx-\bfx^\mathrm{PBH}(t)}{~||\bfx-\bfx^\mathrm{PBH}(t)||^3} \quad.
\label{eq:gravitational_force}
\end{equation}
Given its mass $M^\mathrm{PBH}$ and trajectory (\ref{eq:trajectory}), 
the gravitational force that the PBH exerts on particles at any location within Earth may be calculated.

On Earth's surface $\partial \Omega$, traction-free boundary conditions must be satisfied,
\begin{equation}
\hat \bfn  \cdot \bfT = \mathbf{0} \quad (\bfx\in\partial\Omega,~\forall t) \quad,
\label{eq:free_surface_condition}
\end{equation}
where $\hat\bfn$ denotes the unit outward normal along the boundary, i.e., on the free surface $\partial\Omega$ in this case.
It is critical to note that Earth consists of several major shells: the crust and mantle that behave
as solids on seismological time scales; the outer core that freely flows like a liquid; the inner core that is a dense solid 
comprised of mostly iron. Hence, the boundaries between these layers, namely, the Core-Mantle Boundary (CMB) and the Inner-Core Boundary (ICB),
involve solid-fluid interactions. The boundary conditions on fluid-solid internal boundaries are continuity of normal displacement and traction,
\begin{equation}
\left[\hat\bfn\cdot\,\,\bfs\right]^{+}_{-}=0 \quad (\bfx\in\Sigma,~\forall t) \quad,
\label{eq:displacement_continuity_condition}
\end{equation}
\begin{equation}
\left[\hat\bfn\cdot\bfT\right]^{+}_{-}=\mathbf{0} \quad (\bfx\in\Sigma,~\forall t) \quad,
\label{eq:traction_continuity_condition}
\end{equation}
where $\left[\,g\,\right]^{+}_{-}$ denotes the jump of any function $g$ across an internal boundary $\Sigma$ (either the CMB or ICB).
Together with the initial conditions
\begin{equation}
\bfs(\bfx,0)=\mathbf{0} \quad, \quad  \partial_t\bfs(\bfx,0)=\mathbf{0} \quad,
\label{eq:initial_condition}
\end{equation}
the system of governing equations presented in this section may now be solved numerically.

\subsection{Weak Form}
\label{sec:weak_form}

The wave equation (\ref{eq:governing_equation}) is in differential form, which is generally referred to as the strong formulation.
Classical numerical methods, such as finite-difference and pseudo-spectral methods, usually involve a strong formulation.
However, to account for solid-fluid boundary conditions properly, a weak formulation or integral form of the wave equation
is preferable. The weak formulation is obtained by taking the dot product of equation (\ref{eq:governing_equation}) with an
arbitrary test vector $\bfw$, and integrating over Earth's volume $\Omega$:
\begin{equation}
\label{eq:weak_form}
\int_\Omega~\left( \rho~\partial^2_t \bfs -\bfnabla \cdot \bfT\right) \cdot \bfw~\mathrm{d}^3\bfx
= \int_\Omega~\bff \cdot \bfw ~\mathrm{d}^3\bfx \quad.
\end{equation}

\subsubsection{Crust \& Mantle}

The crust and mantle occupy the solid outer parts of  Earth.
Using the divergence theorem, equation (\ref{eq:weak_form}) in the crust and mantle 
may be simplified as
\begin{equation}
\label{eq:crust_mantle}
\int_{\Omega_\mathrm{M}}~\rho~\partial^2_t \bfs \cdot \bfw~\mathrm{d}^3\bfx 
+ \int_{\Omega_\mathrm{M}}~\bfnabla\bfw:\bfT~\mathrm{d}^3\bfx
- \int_{\partial\Omega}~\hat\bfn \cdot \bfT \cdot \bfw~\mathrm{d}^2\bfx
+ \int_{\mathrm{CMB}}~\hat\bfn \cdot \bfT \cdot \bfw~\mathrm{d}^2\bfx
=\int_{\Omega_\mathrm{M}}~\bff \cdot \bfw ~\mathrm{d}^3\bfx \quad,
\end{equation}
where $\Omega_\mathrm{M}$ denotes the volume of the crust and mantle, $\mathrm{CMB}$ the Core-Mantle Boundary and
$\partial\Omega$ Earth's free surface.

The traction continuity condition on the CMB, i.e., equation (\ref{eq:traction_continuity_condition}), reduces to
\begin{equation}
\hat\bfn\cdot\bfT=-p~\hat\bfn \quad \quad (\bfx\in\mathrm{CMB}, ~\forall t) \quad,
\label{eq:BC_CMB}
\end{equation}
where $p$ is the pressure field in the outer core.

Taking into consideration the free surface boundary condition (\ref{eq:free_surface_condition}) and 
traction continuity condition on the CMB (\ref{eq:BC_CMB}),  equation (\ref{eq:crust_mantle}) may be rewritten as
\begin{equation}
\label{eq:crust_mantle_final}
\int_{\Omega_\mathrm{M}}~\rho~\partial^2_t \bfs \cdot \bfw~\mathrm{d}^3\bfx 
+ \int_{\Omega_\mathrm{M}}~\bfnabla\bfw:\bfT~\mathrm{d}^3\bfx
- \int_{\mathrm{CMB}}~p~\hat\bfn \cdot \bfw~\mathrm{d}^2\bfx
=\int_{\Omega_\mathrm{M}}~\bff \cdot \bfw ~\mathrm{d}^3\bfx \quad.
\end{equation}
It is the pressure field on the CMB that facilitates the exchange of information from the outer core to the mantle.

\subsubsection{Outer Core}

The outer core is a liquid in which the governing equation simplifies significantly.
The displacement field in the outer core $\bfs^\mathrm{fluid}$ may be expressed in terms  of a scalar potential $\chi$
as \citep{Koma02a}
\begin{equation}
\label{eq:fluid_potential}
\bfs^\mathrm{fluid}=\frac{1}{\rho} \bfnabla\chi\quad.
\end{equation}
Substituting this definition into equation (\ref{eq:governing_equation}), we may obtain a scalar governing equation for fluids
\begin{equation}
\label{eq:governing_equation_fluid}
\frac{1}{\kappa}\partial^2_t\chi-\bfnabla\cdot\left(\frac{1}{\rho}\bfnabla\chi\right)
=-\frac{1}{\kappa}\phi \quad.
\end{equation}
The weak form of this equation is
\begin{equation}
\label{eq:outer_core}
\int_{\Omega_\mathrm{O}}~\frac{w}{\kappa}~\partial^2_t\chi~\mathrm{d}^3\bfx
+\int_{\Omega_\mathrm{O}}~\frac{1}{\rho}~\bfnabla w\cdot\bfnabla\chi~\mathrm{d}^3\bfx
-\int_{\mathrm{CMB}}~\frac{w}{\rho}~\bfnabla\chi\cdot\hat\bfn~\mathrm{d}^2\bfx
+\int_{\mathrm{ICB}}~\frac{w}{\rho}~\bfnabla\chi\cdot\hat\bfn~\mathrm{d}^2\bfx
=-\int_{\Omega_\mathrm{O}}~\frac{w}{\kappa}\phi~\mathrm{d}^3\bfx \quad,
\end{equation}
where $\Omega_\mathrm{O}$ denotes the volume of the outer core and $w$ is an 
arbitrary scalar test function.
The continuity condition for the normal component of the displacement field on the CMB and ICB (\ref{eq:displacement_continuity_condition}) reduces to
\begin{equation}
\frac{1}{\rho}\bfnabla\chi\cdot\hat\bfn=\hat\bfn\cdot\bfs^\mathrm{solid} \quad\quad (\bfx\in\mathrm{CMB ~or~ ICB},~\forall t) \quad,
\end{equation}
where $\bfs^\mathrm{solid}$ refers to the displacement on the solid side of the boundary.
Now equation (\ref{eq:outer_core}) may be rewritten as
\begin{equation}
\label{eq:outer_core_final}
\int_{\Omega_\mathrm{O}}~\frac{w}{\kappa}~\partial^2_t\chi~\mathrm{d}^3\bfx
+\int_{\Omega_\mathrm{O}}~\frac{1}{\rho}~\bfnabla w\cdot\bfnabla\chi~\mathrm{d}^3\bfx
-\int_{\mathrm{CMB}}~w~\hat\bfn\cdot\bfs~\mathrm{d}^2\bfx
+\int_{\mathrm{ICB}}~w~\hat\bfn\cdot\bfs~\mathrm{d}^2\bfx
=-\int_{\Omega_\mathrm{O}}~\frac{w}{\kappa}\phi~\mathrm{d}^3\bfx \quad.
\end{equation}
It is the normal component of the displacement field at the CMB \& ICB that passes information from the mantle \& the inner core 
to the outer core, respectively. By definition, the pressure field in the outer core is
\begin{equation}
\label{eq:pressure_definition}
p=\mbox{}-\kappa~\bfnabla\cdot\bfs=\mbox{}-\kappa~\bfnabla\cdot\left(\frac{1}{\rho}\bfnabla\chi\right)=\mbox{}-\partial^2_t\chi-\phi
\quad,
\end{equation}
where in the last equality we have used equation~(\ref{eq:governing_equation_fluid}).

\subsubsection{Inner Core}

The weak form in the solid inner core is similar to the crust and mantle, namely
\begin{equation}
\label{eq:inner_core_final}
\int_{\Omega_\mathrm{I}}~\rho~\partial^2_t \bfs \cdot \bfw~\mathrm{d}^3\bfx 
+ \int_{\Omega_\mathrm{I}}~\bfnabla\bfw:\bfT~\mathrm{d}^3\bfx
+ \int_{\mathrm{ICB}}~p~\hat\bfn \cdot \bfw~\mathrm{d}^2\bfx
=\int_{\Omega_\mathrm{I}}~\bff \cdot \bfw ~\mathrm{d}^3\bfx \quad,
\end{equation}
where $\Omega_\mathrm{I}$ denotes the volume of the inner core.
Equations (\ref{eq:crust_mantle_final}), (\ref{eq:outer_core_final}) and (\ref{eq:inner_core_final}) only hold under simple circumstances,
because they do not take into account rotation or self-gravitation.
General but more complex formulations may be found in \cite{Koma02a,Koma02b};
such complications are accommodated by the spectral-element solver SPECFEM3D\_GLOBE.

\subsection{Black Hole Singularity}

\label{section:black_hole_singularity}

One of the main differences between earthquake and black hole simulations lies in the
fact that we have to deal with different source terms in equations (\ref{eq:crust_mantle_final}), (\ref{eq:outer_core_final}) and (\ref{eq:inner_core_final}), namely
\begin{equation}
\int_{\Omega_\mathrm{O}}~\frac{\phi}{\kappa}~w~\mathrm{d}^3\bfx
\quad \mathrm{and} \quad
\int_{\Omega_\mathrm{M}+\Omega_\mathrm{I}}~\bff \cdot \bfw~\mathrm{d}^3\bfx \quad, \nonumber
\end{equation}
where $\phi$ and $\bff$ are determined by (\ref{eq:gravitational_potential}) and (\ref{eq:gravitational_force}), respectively.
Unlike earthquake simulations, where sources are associated with fault planes, here we encounter volumetric body forces.
When a point within Earth is close to the location of the black hole, the gravitational force and potential become singular.
To avoid such unphysical singularities, we implement the source term as
\begin{equation}
\begin{split}
\widetilde\bff(\bfx,t)=
       \begin{dcases}
       \mbox{}-\,{G\,M^\mathrm{PBH}\,\rho(\bfx)}\,\frac{\bfx-\bfx^\mathrm{PBH}(t)}{~||\bfx-\bfx^\mathrm{PBH}(t)||^3} \quad,\qquad\qquad &||\bfx-\bfx^\mathrm{PBH}(t)|| \geq R_\mathrm{crit} \quad, \\
       \mbox{}-\,{G\,M^\mathrm{PBH}\,\rho(\bfx)}\,\frac{\bfx-\bfx^\mathrm{PBH}(t)}{R_\mathrm{crit}^3} \quad,\qquad\qquad &||\bfx-\bfx^\mathrm{PBH}(t)|| < R_\mathrm{crit} \quad,
        \end{dcases}
\label{eq:force_implementation1}
\end{split}
\end{equation}

\begin{equation}
\begin{split}
\widetilde\phi(\bfx,t)=
                   \begin{dcases}
                   \mbox{}-\,\frac{G\,M^\mathrm{PBH}}{~||\bfx-\bfx^\mathrm{PBH}(t)||} \quad, &||\bfx-\bfx^\mathrm{PBH}(t)|| \geq R_\mathrm{crit}\quad,\qquad \\
                   \mbox{}-\frac{3~G\,M^\mathrm{PBH}}{2~R_\mathrm{crit}}
                   +\,\frac{G\,M^\mathrm{PBH}}{2~R_\mathrm{crit}^3} ||\bfx-\bfx^\mathrm{PBH}(t)||^2 \quad,\qquad  &||\bfx-\bfx^\mathrm{PBH}(t)||<R_\mathrm{crit} \quad,
                   \end{dcases}
\label{eq:force_implementation2}
\end{split}
\end{equation}
where $R_\mathrm{crit}$ is a chosen critical length scale within which the amplitude of the gravitational force
decreases linearly. 
It may be shown mathematically that $\widetilde \phi$ and $\widetilde \bff$ capture all features of the original source terms $\phi$ and $\bff$, as long as the critical length is less than the size of the numerical grid.
In our simulations, the size of an element is $\sim 30$ km, while
$R_\mathrm{crit}$ is around $\sim 6$ m, i.e., five thousand times smaller than the element size. 
Convergence tests using different $R_\mathrm{crit}$ illustrate that further decreasing $R_\mathrm{crit}$ does not change
the simulation results (Appendix~\ref{appendix:convergent_tests}).

The last, but equally important issue is the selection of the time step for the numerical simulations. 
As mentioned previously, the speed of the black hole is typically a hundred times larger than the seismic wave speeds, and
it takes only a few minutes for the PBH to traverse Earth.
Therefore, we need a much smaller time step than usual to resolve the trajectory of the black hole.
Fig.~\ref{fig:resolve_trajectory} in Appendix~\ref{appendix:convergent_tests} explains why this is important.
In practice, we choose a time step that is one thousandth of the original time step when the PBH is close to Earth.

\section{Results of Spectral-Element Simulations}\label{section:results}

\begin{figure}
\centering
\includegraphics[width=0.50\textwidth,angle=0]{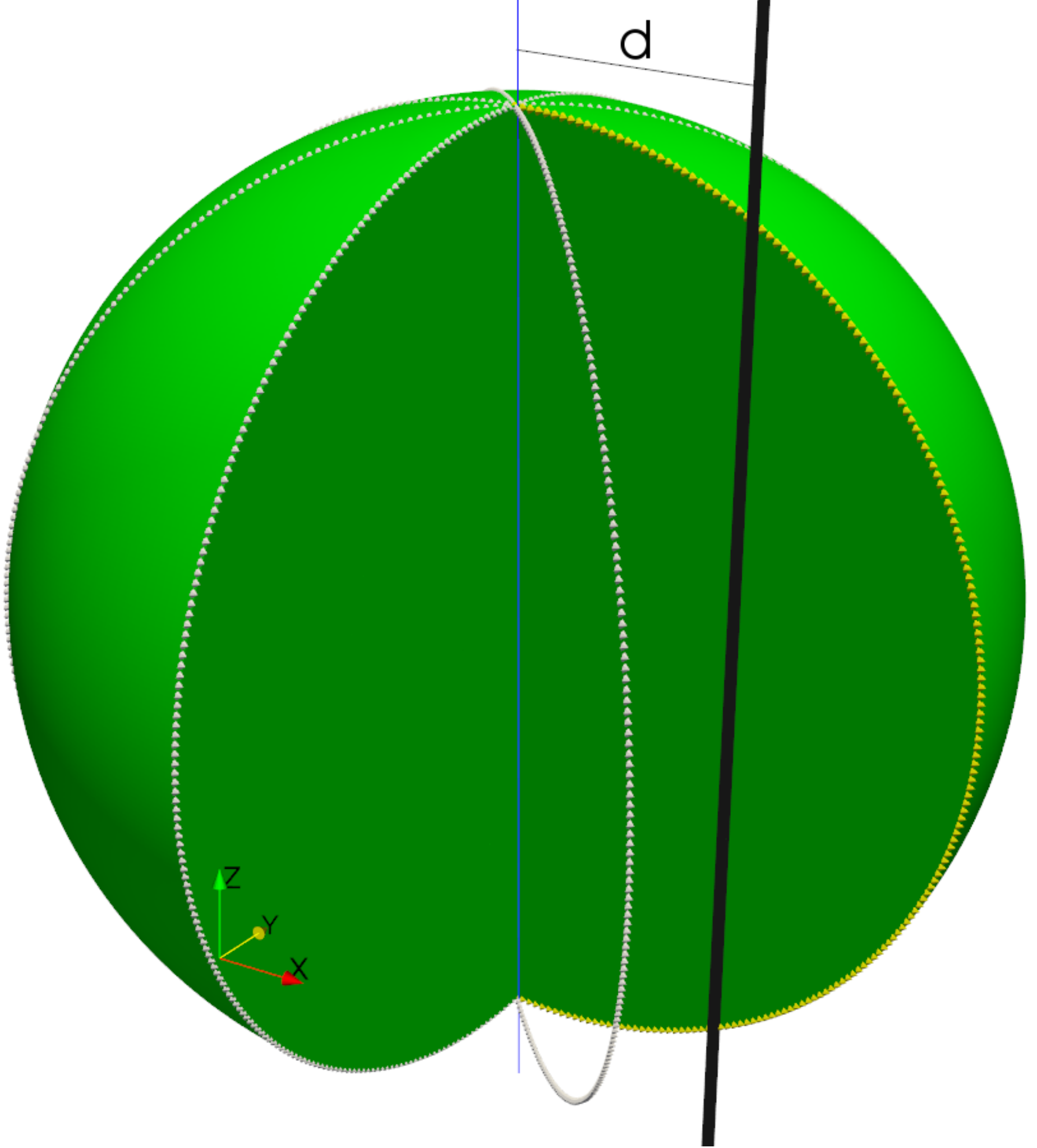}
\caption[fig:set_up]
{
Set-up for the numerical simulations.
The thick black line denotes the PBH trajectory, where the distance between
the trajectory and the parallel diameter of the Earth is denoted by $d$. 
For example, when the PBH passes through the center of Earth, $d=0$;
when the PBH misses the Earth, $d>R_\oplus$, where $R_\oplus=6371$~km is the radius of Earth.
Hypothetical seismic stations denoted by triangles are deployed around the globe,
but only the most significant ones (yellow triangles) will be shown later.
}
\label{fig:set_up}
\end{figure}

The set-up of our simulations is illustrated in Fig.~\ref{fig:set_up}.
Different trajectories as well as two distinct black hole speeds, namely, 200~km~s$^{-1}$ and 400~km~s$^{-1}$, are considered.
The simulations are carried out on a cubed-sphere mesh \citep{Koma02a,Koma02b} comprising 8.2 million elements.
Given the wave-speed structure of Earth, the shortest period the mesh can resolve is around 15~s.
We set the black hole mass $M^\mathrm{PBH} = 10^{15}$~g,
noting that amplitudes of the resultant seismic waves scale linearly with $M^\mathrm{PBH}$.

\begin{figure}
\centering
\includegraphics[width=0.90\textwidth,angle=0]{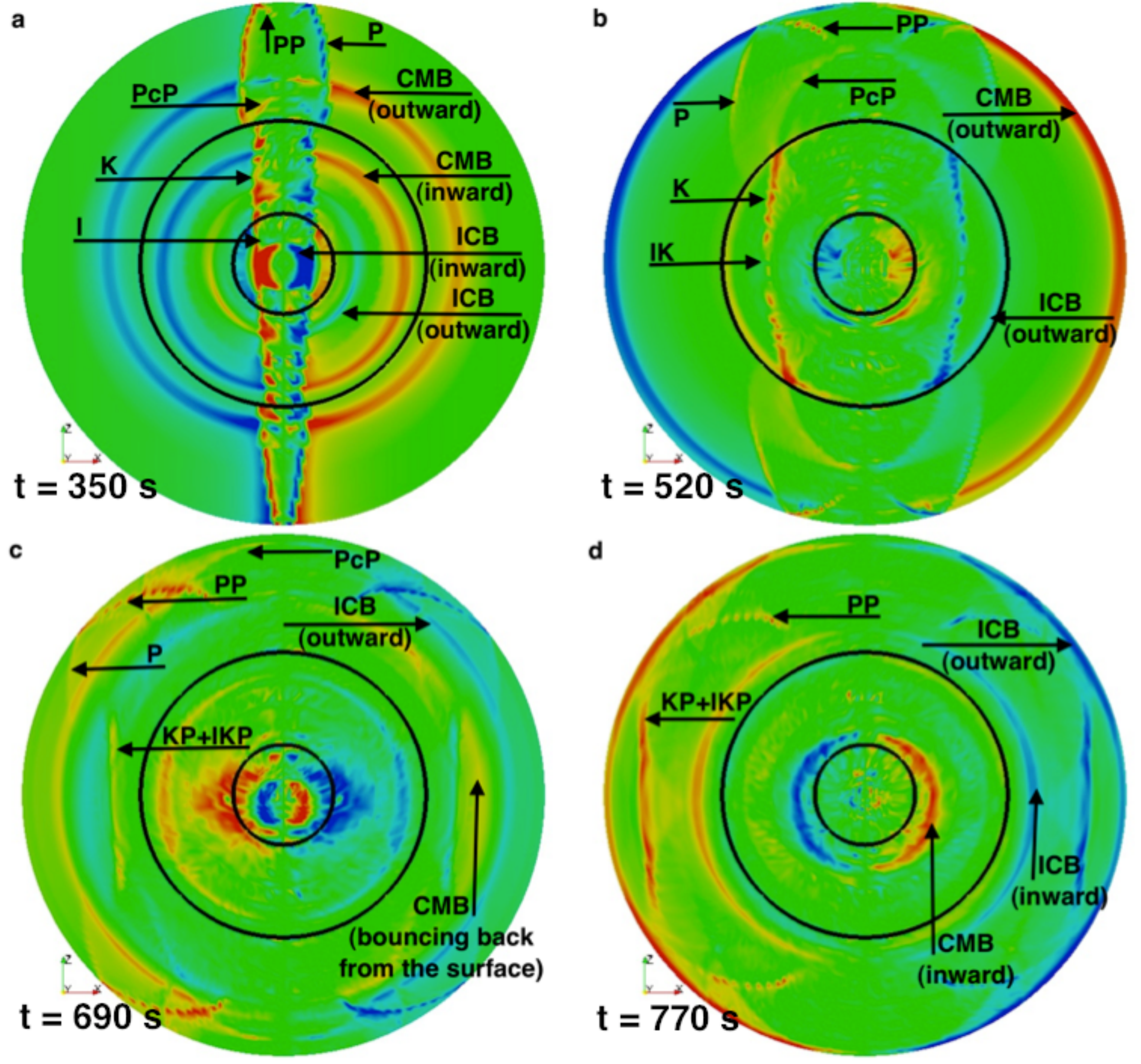}
\caption[fig:snapshots]
{
Snapshots at four different instantaneous times.
Shown are particle motions along the $x$-direction in the $x$-$z$ plane through Earth's center.
Red represents motion to the left, i.e., the negative $x$ direction, whereas blue indicates that particles are moving to the right.
\textbf{a} Generation of cylindrical and spherical waves. The PBH has entered Earth from the top and is about to
exit from the opposite side.
Some wave groups are labeled using commonly used seismological nomenclature, and may be viewed together with those shown in Fig.~3. 
\textbf{b} Propagation of cylindrical and spherical waves.
Spherical waves from the CMB reach Earth's surface everywhere at the same time, leaving the linear features we see in Fig.~3. 
Because of the non-homogeneous seismic wave speeds, cylindrical waves are bent and gradually lose their cylindrical shape.
\textbf{c} Spherical waves bounce back from the surface,
whereas cylindrical waves continue to arrive at stations from high latitudes to the equator.
\textbf{d} Spherical waves from the ICB reach the surface and cylindrical waves
generated in the outer core \& inner core (KP/IKP) are about to hit the equator.
It is important to understand that these snapshots are for illustrative purposes only.
A large critical length scale is used to eliminate high-frequency numerical ``noise''
and hence the amplitudes of cylindrical waves are greatly underestimated compared to spherical waves.
}
\label{fig:snapshots}
\end{figure}

\begin{figure}
\centering
\includegraphics[width=0.98\textwidth,angle=0]{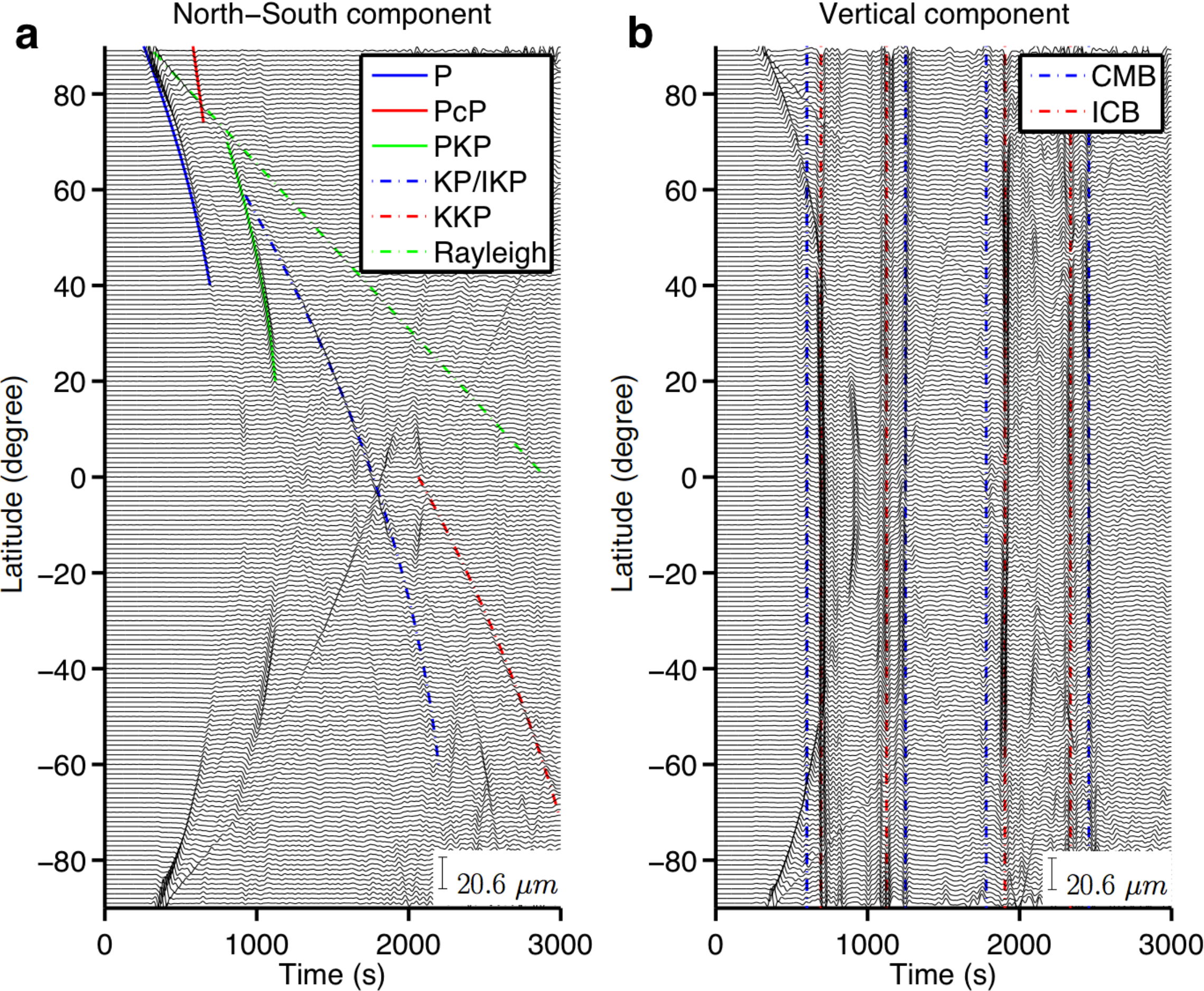}
\caption[fig:trajectory1]
{
Simulation results for $v=200$~km~s$^{-1}$ and $d=0$ in Fig.~1. 
East-West components are omitted since no signals can be observed due to symmetry.
Snapshots in Fig.~2 
may help to better understand phase labels.
\textbf{a} North-South components of ground motion.
The earliest arrival (solid blue line) is often called the Mach wave, similar to an earthquake P phase.
The solid red line represents seismic waves reflected at the CMB, corresponding to an earthquake PcP phase.
The solid green line highlights the PKP phase traveling through the outer core.
Phases KP and IKP are represented by the dashed blue line, generated either in the outer or inner core, respectively.
The KKP phase, denoted by the dashed red line, is reflected once at the CMB before entering the mantle.
The phase denoted by the dashed green line has a wave speed of $\sim 3.8~\mathrm{km~s}^{-1}$, i.e., the Rayleigh surface wave.
\textbf{b} Vertical components of ground motion.
Besides imprints of the same phases seen on the North-South components,
phases emerge that arrive at all stations simultaneously.
These involve spherical boundary waves generated at the CMB \& ICB, labeled by dashed blue and red lines, respectively.
Time intervals between these phases correspond to reverberations.
For example, the time interval between the first and second blue dashed lines corresponds to waves traveling
across the outer and inner cores through the Earth's center,
while the time interval between the first and third blue dashed lines corresponds to waves traveling
across the entire planet.
}
\label{fig:trajectory1_200}
\end{figure}

An instructive case is that of a PBH whose trajectory coincides with the center of Earth, i.e., $d = 0$ in Fig.~\ref{fig:set_up}.
We use the 1-D Preliminary Reference Earth Model \citep{PREM},
aligning the PBH trajectory with Earth's rotation axis.
Four snapshots from this simulation are shown in Fig.~\ref{fig:snapshots},
illustrating generation, propagation, transmission and reflection of PBH-induced seismic waves.
Seismograms from this simulation are displayed in Fig.~\ref{fig:trajectory1_200}.
Due to symmetry, East-West ground motions are very weak compared to the other two components,
and are thus not shown.
The North-South component shows not only distinctive arrivals due to cylindrical waves from the supersonic motion of the PBH,
but also classical Rayleigh surface waves and reflected/transmitted core phases.
The vertical component, besides containing signatures similar to the North-South component,
registers unique waves that arrive at almost the same time everywhere on Earth's surface,
distinct from those arising due to earthquakes.
These waves are generated at the Core-Mantle Boundary (CMB) and Inner-Core Boundary (ICB) due to free-slip interactions between the solid mantle \& inner core and the
liquid outer core.

We study four additional PBH trajectories, two of which pass through Earth's interior
($d=0.4~R_\oplus$ and $d=0.8~R_\oplus$) while the other two pass nearby ($d=1.2~R_\oplus$ and $d=1.6~R_\oplus$).
A faster PBH traveling at a speed of 400~km~s$^{-1}$ is also considered.
To avoid redundancy, only $d=0.8~R_\oplus$ for the faster PBH is shown in Fig.~\ref{fig:trajectory3_400}
(more images may be found in Appendix~\ref{appendix:more_examples}).
On the North-South component, clearly visible are direct arrivals due to the supersonic motion of the source
as well as Rayleigh surface waves, which behave in the same manner as those shown in Fig.~\ref{fig:trajectory1_200}.
Distinctions from the simple example also emerge. For example, the entrance and exit points of the PBH
change from the poles to latitudes around $\pm40^\circ$, as indicated by the focus of the surface waves.
Waves arriving after the surface waves are coherent but more difficult to interpret.
The vertical component is largely unchanged from the simple example.
Besides other signatures found on the North-South component, linear features due to the spherical waves exist as well.
For cases where the PBH does not hit our planet, we see no coherent signals on the horizontal components.
But spherical waves are still visible on the vertical component, since they are gravitationally induced at the CMB and ICB.
Simulations in 3-D Earth model S40RTS \citep{S40RTS} preserve these key features, as illustrated in Appendix~\ref{appendix:more_examples}.

\begin{figure}
\centering
\includegraphics[width=0.98\textwidth,angle=0]{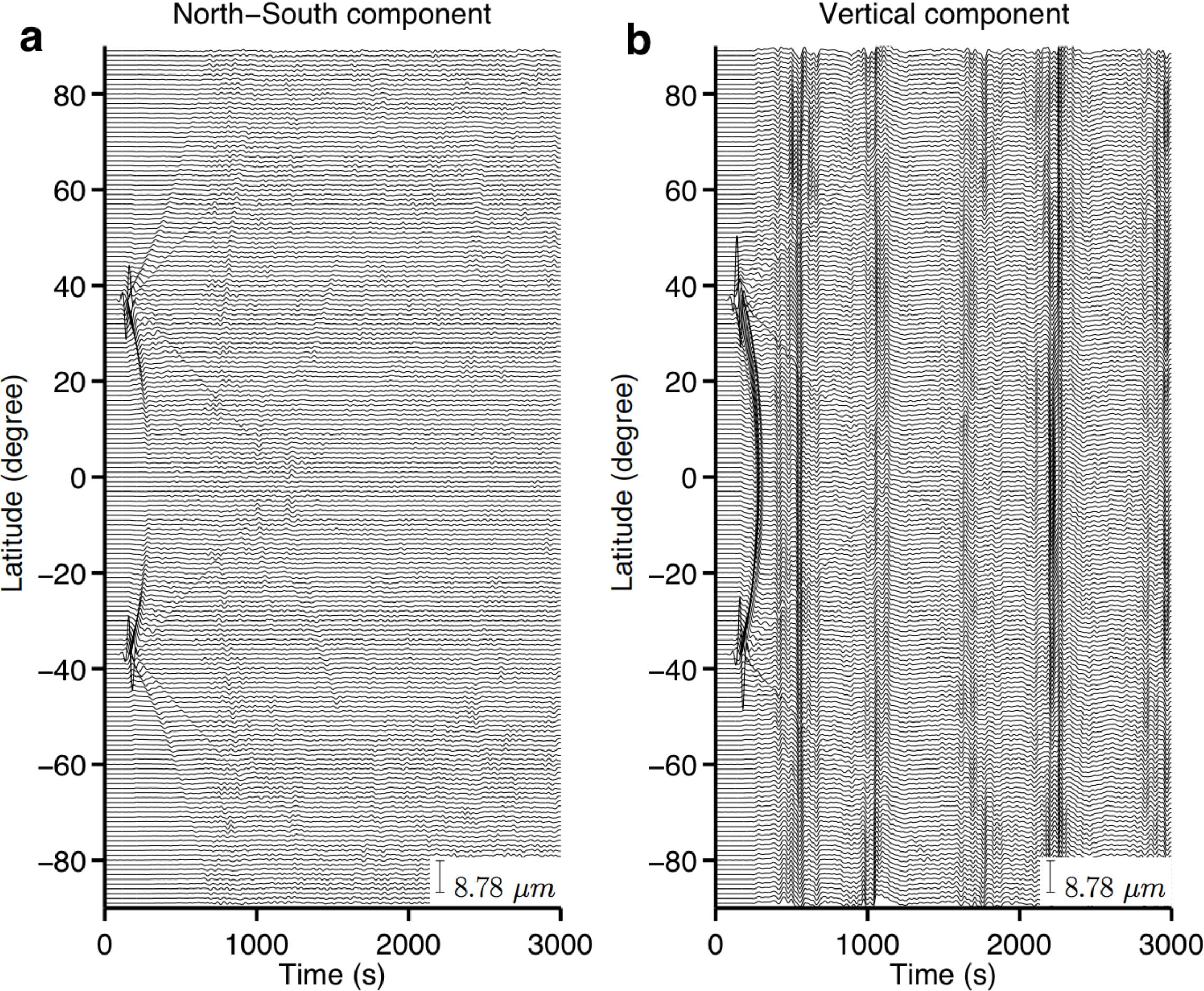}
\caption[fig:trajectory3]
{
Simulation results for $d=0.8~R_\oplus$ and $v=400~\mathrm{km~s}^{-1}$.
\textbf{a} North-South component ground motion.
\textbf{b} Vertical-component ground motion.
Both cylindrical and spherical waves can be seen clearly.
Focusing points of the Rayleigh surface waves move to latitudes of $\pm 40^\circ$, reflecting changes
in the entrance and exit locations of the PBH.
}
\label{fig:trajectory3_400}
\end{figure}

For all cases, ground motions are on the order of $10^{-6}$~m.
For comparison, ground motions due to the Bolivia $M_\mathrm{w}=8.1$ earthquake on June 9, 1994 are on the order of $10^{-3}$~m.
From an energy perspective, seismic energy injected by the $10^{15}$~g PBH is six orders in magnitude lower than that of the Bolivia earthquake,
corresponding to an earthquake of magnitude $M_\mathrm{w}=4$.

\clearpage
\section{Normal-Mode Excitation}
\label{section:modes}

\begin{figure}
\centering
\includegraphics[width=0.70\textwidth,angle=0]{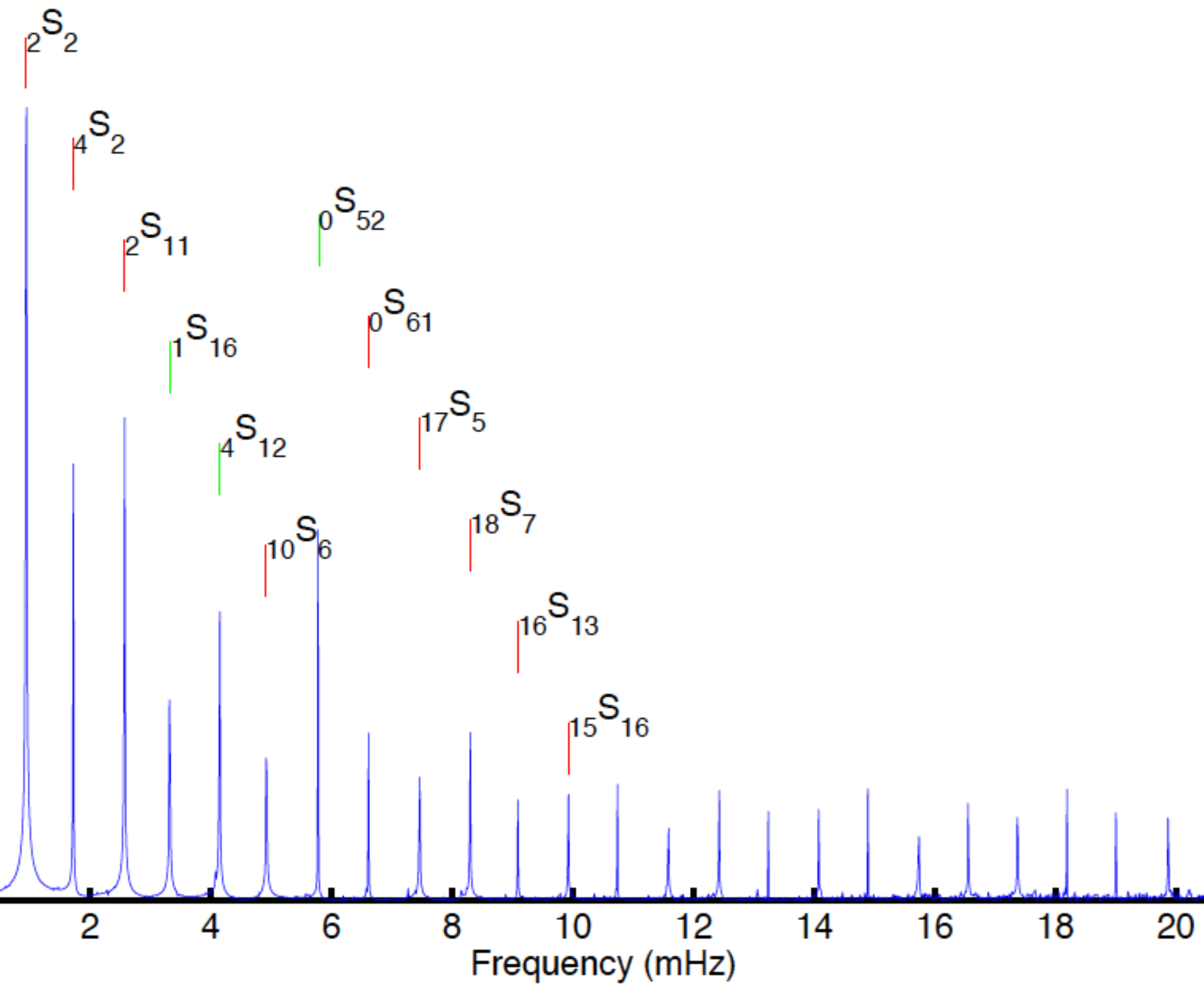}
\caption[fig:spectra_vertical]
{
Stacked spectra of vertical-component seismograms at all stations.
Long time series of 20~hours are used to calculate the spectra,
instead of the 50-minute seismograms shown in Fig.~3. 
Labeled normal modes are from a mode calculation in 1-D model PREM,
where red lines denote modes that are not commonly observed, such as $_2S_2$.
Differences between the peaks and the labeled modes are $\sim$ 0.02~mHz,
around the sampling frequency resolved by a 20-hour long record.
Besides these unusual modes, the evenly-spaced peaks define another distinct PBH feature.
The behavior of these modes is similar to radial modes, $_nS_0$, and their frequency spacing corresponds to
the compressional wave travel time at an epicentral distance of $180^\circ$ (PKIKP).
These modes correspond to the spherical boundary waves we see in Fig.~3. 
}
\label{fig:spectra_vertical}
\end{figure}

Another distinctive feature of PBH-induced seismic waves is the appearance of unusual spheroidal normal modes (which are not excited by earthquakes)
in spectra of the time series, as shown in Fig.~\ref{fig:spectra_vertical}.

The significance of the power spectra arises from normal-mode analysis \citep{DT}.
Any motions on Earth can be decomposed into a set of periodic oscillations --- Earth's free oscillations or normal modes:
\begin{equation}
\bfs(\bfx,t)=\mathrm{Re}\sum_k~(i\omega_k)^{-1}~\bfs_k(\bfx)~\int_{-\infty}^t~C_k(t')~\mathrm{exp}[i\omega_k(t-t')]~\rmd t' \quad,
\label{eq:normal_modes_general}
\end{equation}
where $\bfs_k$ denotes the $k$th mode, $\omega_k$ its eigenfrequency and $C_k$ its excitation.

\subsection{Earthquake Excitation}

For an earthquake, the excitation term $C_k$ is controlled by the location of the earthquake and its mechanism:
\begin{equation}
C_k(t)=\int_{\Sigma^t}~\bfm(\bfx,t)~{\bm:}~{\bm\varepsilon}_k^*(\bfx)~\rmd^2\bfx \quad,
\end{equation}
where $\Sigma^t$ represents the fault plane at time $t$, $\bfm$ the moment density tensor and
${\bm\varepsilon}_k$ the strain tensor associated with the $k$th mode
\begin{equation}
{\bm\varepsilon}_k=\frac{1}{2}\left[\bfnabla\bfs_k+\left(\bfnabla\bfs_k\right)^T\right] \quad.
\end{equation}
When the point source approximation applies, as is often the case in global seismology, the excitation term
reduces to
\begin{equation}
C_k(t)\approx\bfM(t)~{\bm:}~{\bm\varepsilon}_k^*(\bfx_s)\quad,
\label{eq:earthquake_excitation}
\end{equation}
where the strain tensor ${\bm\varepsilon}_k$ at the location of the point source $\bfx_s$ interacts with the moment tensor
\begin{equation}
\bfM(t)=\int_{\Sigma^t}~\bfm(\bfx,t)~\rmd^2\bfx \quad.
\end{equation}
Equation (\ref{eq:earthquake_excitation}) determines not only which modes can be excited by an earthquake, but also how strongly excited these modes are.
It is critical to note that earthquakes only occur in the crust \& upper mantle. Therefore, modes that are not
sensitive to the shallow parts of Earth cannot be excited by natural earthquakes;
examples include the Slichter mode (${}_1S_1$) and Stoneley modes trapped at the CMB \& ICB.

\subsection{PBH Excitation}
\label{appendix:PBH_excitation}

A PBH triggers normal modes in a different way than earthquakes.
The excitation follows the form of a distributed force
\begin{equation}
C_k(t)=\int_\Omega~\bff(\bfx,t)\cdot\bfs_k^*(\bfx)~\rmd^3\bfx \quad,
\label{eq:PBH_excitation}
\end{equation}
where $\bff$ satisfies
\begin{equation}
\bff(\bfx,t)=G~M^\mathrm{PBH}~\rho(\bfx)~\bfnabla\frac{1}{~||\bfx-\bfx^\mathrm{PBH}(t)||} \quad.
\label{eq_appendix:PBH_force}
\end{equation}
Substituting equation (\ref{eq_appendix:PBH_force}) into (\ref{eq:PBH_excitation}), we obtain
\begin{eqnarray}
C_k(t)
&=&\int_\Omega~G~M^\mathrm{PBH}~\rho(\bfx)~\bfs_k^*(\bfx)\cdot\bfnabla
        \left[\frac{1}{r'}\sum_{l=0}^{\infty}\left(\frac{4\pi}{2l+1}\right)\left(\frac{r}{r'}\right)^l\sum_{m=-l}^{l}Y_{lm}(\theta,\phi)~Y_{lm}^*(\theta',\phi')\right]~\rmd^3\bfx\nonumber\\
 &=&\int_\Omega~G~M^\mathrm{PBH}~\rho(\bfx)~\bfs_k^*(\bfx)\cdot\left(\hat\bfr~\partial_{r}+r^{-1}\bfnabla_1\right)
         \left[\frac{1}{r'}\sum_{l=0}^{\infty}\left(\frac{4\pi}{2l+1}\right)\left(\frac{r}{r'}\right)^l\sum_{m=-l}^{l}Y_{lm}(\theta,\phi)~Y_{lm}^*(\theta',\phi')\right]~\rmd^3\bfx \quad,\nonumber\\
\label{eq_appendix:temp}
\end{eqnarray}
where $r\equiv||\bfx||$, $r'\equiv||\bfx^\mathrm{PBH}||$,
$\bfnabla_1\equiv\hat{\bm\theta}~\partial_\theta+\hat{\bm\phi}~(\sin\theta)^{-1}\partial_\phi$ denotes the surface gradient on the unit sphere,
and $Y_{lm}(\theta,\phi)$ are spherical harmonics \citep{Edmonds60}.
Note that this expansion is valid only for $r'>r$. When the PBH is within Earth, $r$ may be larger than $r'$ for some points $\bfx$.
In that case, we need to interchange $r$ and $r'$, leading to a more complex result than the one considered here.

Mode $\bfs_k$ may be a spheroidal mode $^S\bfs_k$ or a toroidal mode $^T\bfs_k$:
\begin{eqnarray}
^S\bfs_k(\bfx)&=&{}_nU_l(r)~Y_{lm}(\theta,\phi)~\hat\bfr+\frac{1}{\sqrt{l(l+1)}}~{}_nV_l(r)~\bfnabla_1Y_{lm}(\theta,\phi) \quad,\\
^T\bfs_k(\bfx)&=&\frac{\mbox{}-1}{\sqrt{l(l+1)}}~{}_nW_l(r)~\left(\hat\bfr\times\bfnabla_1\right)Y_{lm}(\theta,\phi) \quad,
\end{eqnarray}
where $l$, $m$ and $n$ are the degree, order and overtone number of the mode, respectively; ${}_nU_l$, ${}_nV_l$ and ${}_nW_l$
are the corresponding radial eigenfunctions.
Due to the relations
\begin{eqnarray}
\hat\bfr\cdot\bfnabla_1&=&0 \quad,\\
\hat\bfr\cdot\left(\hat\bfr\times\bfnabla_1\right)&=&0 \quad,\\
\bfnabla_1\cdot\left(\hat\bfr\times\bfnabla_1\right)&=&0 \quad,
\end{eqnarray}
it is obvious that no toroidal modes can be generated by a PBH, because the excitation term vanishes.
For spheroidal modes, equation (\ref{eq_appendix:temp}) reduces to
\begin{eqnarray}
\label{eq_appedix:PBH_excitation_spheroidal1}
C_k(t)&=&~{}_nA_l(t)~Y^*_{lm}\left(~\theta'(t),\phi'(t)~\right) \quad,\\
{}_nA_l(t)&=&\frac{4\pi~G~M^\mathrm{PBH}}{(2l+1)~\left(r'(t)\right)^{l+1}}
            ~\int_0^{R_\oplus}~\rho(r)~r^{l+1}~\left[l~{}_nU_l(r) +\sqrt{l(l+1)}~{}_nV_l(r)\right]~\rmd r \quad,
\label{eq_appedix:PBH_excitation_spheroidal2}
\end{eqnarray}
where we have used the orthogonality of spherical harmonics.
Note that the PBH location $\bfx^\mathrm{PBH}$ varies with time,
and therefore $r'$, $\theta'$ and $\phi'$ are also functions of time.
Equations (\ref{eq_appedix:PBH_excitation_spheroidal1}) and (\ref{eq_appedix:PBH_excitation_spheroidal2})
differ from equation (\ref{eq:earthquake_excitation}) in that
the eigenfunctions everywhere along Earth's radius are involved.
The implication is that even modes whose sensitivities are restricted to the deeper parts of Earth,
i.e., modes that cannot be triggered by earthquakes, may be excited by a PBH. 

Substituting equations (\ref{eq_appedix:PBH_excitation_spheroidal1}) and (\ref{eq_appedix:PBH_excitation_spheroidal2})
into equation (\ref{eq:normal_modes_general}), we obtain
\begin{equation}
\bfs(\bfx,t)=\mathrm{Re}\sum_k~(i\omega_k)^{-1}~^S\bfs_k(\bfx)
    ~\int_{-\infty}^t~{}_nA_l(t')~Y^*_{lm}\left(~\theta'(t'),\phi'(t')~\right)~\mathrm{exp}[i\omega_k(t-t')]~\rmd t' \quad.
\label{eq:PBH_s}
\end{equation}
Note again that the summation over $k$ actually denotes triple summations over $l$, $m$ and $n$.
In order to evaluate equation (\ref{eq:PBH_s}) efficiently, the summation over $m$ is carried out analytically using
the spherical-harmonic addition theorem
\begin{equation}
\sum_{m=-l}^{l}~Y^*_{lm}({\theta',\phi'})~Y_{lm}(\theta,\phi)=\left(\frac{2l+1}{4\pi}\right)~P_l(\cos\Theta) \quad,
\end{equation}
where $\cos\Theta=\cos\theta\cos\theta'+\sin\theta\sin\theta'\cos(\phi-\phi')$.
Finally, equation (\ref{eq:PBH_s}) reduces to
\begin{eqnarray}
\bfs(\bfx,t)&=&\left(\frac{2l+1}{4\pi}\right)\mathrm{Re}\sum_{l,n}~(i~{}_n\omega_l)^{-1}
       ~\left[{}_nU_l(r)~\hat\bfr
       ~\int_{-\infty}^t~{}_nA_l(t')~P_{l}\left(\Theta(t')\right)~\mathrm{exp}(-i~{}_n\omega_l~t')~\rmd t'\right.  \nonumber\\
       &+&
       \left. \frac{1}{\sqrt{l(l+1)}}~{}_nV_l(r)
       ~\int_{-\infty}^t~{}_nA_l(t')~\bfnabla_1 P_{l}\left(\Theta(t')\right)~\mathrm{exp}(-i~{}_n\omega_l~t')~\rmd t' \right] ~\mathrm{exp}(i~{}_n\omega_l~t) \quad.
\label{eq:PBH_s_final}
\end{eqnarray}

\begin{figure}[h]
\centering
\includegraphics[width=0.70\textwidth,angle=0]{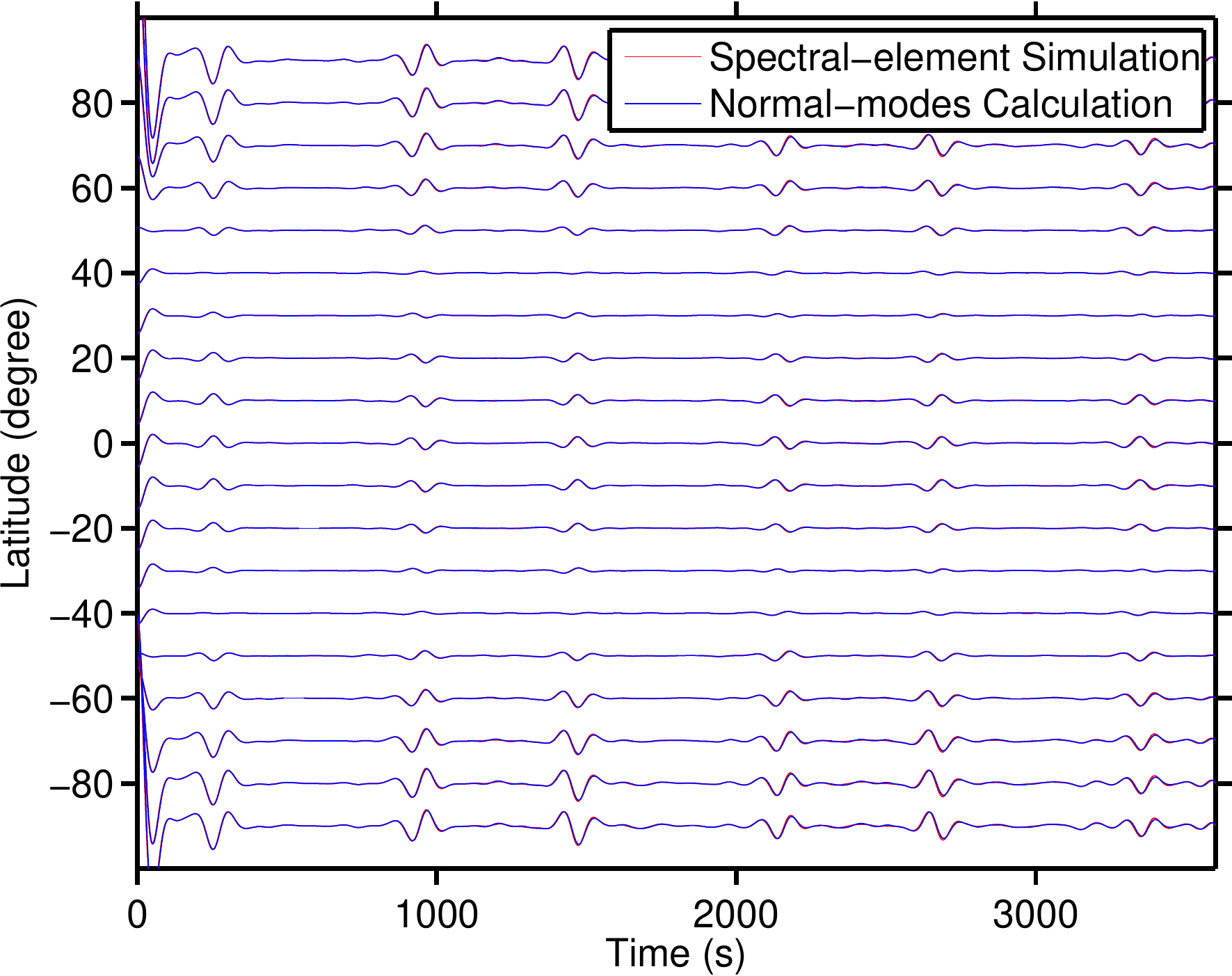}\\
\caption[fig:SEM_NM]
{Benchmark between a normal-mode calculation and a spectral-element simulation
for two stationary, transient PBHs placed symmetrically about Earth's center.
This benchmark confirms the accurate implementation of the volumetric PBH source in the spectral-element code.
}
\label{fig:SEM_NM}
\end{figure}

Equation (\ref{eq:PBH_s_final}) is difficult to implement for a moving PBH, however,
it is straightforward to consider a stationary and transient PBH effect.
In this case, instead of equation (\ref{eq_appendix:PBH_force}), the force term is
\begin{equation}
\bff(\bfx,t)=G~M^\mathrm{PBH}~\rho(\bfx)~\bfnabla\frac{1}{~||\bfx-\bfx_0^\mathrm{PBH} ||}~\delta'(t) \quad,
\end{equation}
where $\delta'(t)$ --- rather than $\delta(t)$ --- is used to avoid inducing a permanent displacement.
In this simpler case, where $r'$, $\theta'$ and $\phi'$ are time-independent,
equation (\ref{eq:PBH_s_final}) takes the following form
\begin{eqnarray}
\bfs(\bfx,t)=\left(-\frac{2l+1}{4\pi}\right)\mathrm{Re}\sum_{l,n}~
     ~{}_nA_l~\left[{}_nU_l(r)~P_{l}\left(\cos\Theta\right)~\hat\bfr+\frac{1}{\sqrt{l(l+1)}}~{}_nV_l(r)~\bfnabla_1 P_{l}\left(\cos\Theta\right)\right]
     ~\mathrm{exp}(i~{}_n\omega_l~t) \quad.
\label{eq:PBH_s_simple}
\end{eqnarray}
We shall use this configuration solely for benchmarking purposes,
testing the generation of spherical waves
by comparing the results of a normal mode calculation with those obtained based on a SEM calculation.
For earthquake sources and impacts, the SEM has already been benchmarked extensively against normal modes
\citep[e.g.,][]{Koma02a,Mesetal2011}.
Figure~\ref{fig:SEM_NM} shows a comparison between a normal-mode calculation and a spectral-element simulation for two stationary and transient PBHs, which are symmetrically placed with respect to Earth's
center. Using two PBHs instead of one ensures that Earth's center remains stationary.

\section{Estimated Event Rates and Non-seismic Effects}\label{section:estimates}

In this section we provide some order-of-magnitude estimates of event rates
and non-seismic effects of a black hole (BH) or quark nugget (QN) moving
through the earth. The radius of a BH is
\begin{equation}
r_\mathrm{BH}\approx 1.5 \times 10^{-15} {\rm m}\ \overline{M}
\quad,
\end{equation}
where $\overline{M}\equiv M/10^{15}$~g is its mass in units of $10^{15}$~g. 
Assuming a QN is a uniform sphere with density
$\rho_\mathrm{QN} = 10^{21}\rm{g}/\rm{m}^3 \rho_0$, with $\rho_0$ of order
unity~\citep{witten84}, it has a radius of
\begin{equation}
r_\mathrm{QN}\approx 6 \times 10^{-3} \rm{m} \left(\,\frac{\overline{M}}{\rho_0}\,\right)^{1/3}
\quad.
\end{equation}
Restricting the discussion to masses less than $\sim 10^{-3} M_\oplus$,
one can see that though
the QN is significantly larger than the BH, as gravitational sources both can still be 
considered point-particles to good approximation. 
Assuming that a fraction $\chi$ of the estimated dark matter density
in the vicinity of the solar system --- see for example~\cite{lisanti_spergel} ---
is in the form of compact objects of mass $\overline{M}$, the expected event rate
is
\begin{equation}
N\approx 4\times10^{-7} {\rm yr}^{-1} \>\frac{\chi\ \bar{v}\ \bar{d}^{\,2}}{\overline{M}}
\quad,
\end{equation}
where $\bar{v}\equiv v/200$~km/s is the velocity in units of $200$~km/s, and
$\bar{d}=b_{\max}/R_\oplus$ is the maximum distance, in units of $R_\oplus$,
out to which the passage of a compact object with impact parameter $b_{\max}$ will
create the kind of impulsive gravitational response that gives rise to the unique
seismic signatures discussed here. A rough estimate suggests $\bar{d}\approx10-20$.
These events are thus quite rare for masses that could produce
observable seismic signatures. However, as we will show below, on the lower
mass end of BHs that are seismically observable, the corresponding non-seismic signatures
will be negligible, and seismology may offer a unique way to observe them.
This is not the case for QNs, which, because of their larger radii, could produce
strong atmospheric explosions.

The Bondi radius $R_B$ for accretion of material onto a compact object is
\begin{equation}
R_B \approx 3\times10^{-9} {\rm m} \> \frac{\overline{M}}{\bar{v}^2}
\quad,
\end{equation}
For the masses considered here, this is negligibly small for BHs, and is inside the radius of a 
QN, thus for QNs the relevant radius is simply $r_\mathrm{QN}$.
Following~\cite{burns_greenstein_verosub}, the radius $R_m$ out to which the gravitational
impulse of the passing compact object could melt surface rock is
\begin{equation}
R_m \approx 5\times 10^{-7} {\rm m}\> \frac{\overline{M}}{\bar{v}}
\quad,
\end{equation}
and following~\cite{labun_birrel_rafelski}, the transverse distance $L$ out
to which rock will be shattered due to the gravitational tidal field is
\begin{equation}
L \approx 10^{-5} {\rm m} \ \overline{M}
\quad.
\end{equation}
Again, both $R_m$ and $L$ are within the radius of the QN, and negligible
for smaller mass BHs. Applying the results of~\cite{ruderman_spiegel},
we may estimate the energy that will be deposited into the atmosphere
by the passage of the compact object as it passes through. For BHs
the main effect is due to the impulsive gravitational acceleration
of air molecules, giving
\begin{equation}
\Delta E^\mathrm{atm}_\mathrm{BH} \approx 0.3 \ {\rm J} \ \left(\,\frac{\overline{M}}{\bar{v}}\,\right)^2
\quad.
\end{equation}
To create an explosion such as the Tunguska event of $\sim 10^{15}-10^{17}$~J, for example,
would require a BH of mass $~10^{8}-10^{9}\, \overline{M}$. QNs cause a similar gravitational
impulse, however their larger surfaces experience a drag that can
deposit significantly larger amounts of energy compared to BHs for the smaller
mass objects:
\begin{equation}
\Delta E^\mathrm{atm}_\mathrm{QN} \approx 4 \times 10^{10} \ {\rm J} \ \bar{v}^2 \left(\,\frac{\overline{M}}{\rho_0}\,\right)^{2/3}
\quad.
\end{equation}
Again, to create Tunguska-like explosions requires massive QNs, however,
in contrast to BHs, smaller QNs can still produce sizable explosions, e.g., a $\overline{M}=1$
QN imparts the energy equivalent of $\sim 10$~tons of TNT. 

The event rates above are exceedingly low for massive ($\overline{M}\gg 1$) compact
objects, though as discussed in~\cite{labun_birrel_rafelski}, direct
impacts may leave features distinguishable from asteroid impacts,
and hence the effective baseline for observation can be extended by
looking at geological features on Earth, Moon, Mercury and Mars. Also,
as BH and QN detectors the gas giants have larger collecting areas, and may 
be useful in constraining compact object dark matter candidates in regimes
not readily accessible by other observations.

\section{Recipe for Detection}
\label{section:recipe}
Based on numerical simulations, we find that PBH transits through Earth generate unique and significant seismic signals.
Our theoretical results may be used to develop a strategy for detecting PBHs.
Unlike earlier suggestions \citep{Davidetal2003}, we propose to use spherical rather than cylindrical waves
as the key detection mechanism. There are several advantages to this approach. First, detection is more reliable
because the simultaneous global arrivals induced by slip at the CMB and ICB are unique and cannot be produced by earthquakes.
In contrast, cylindrical-wave arrivals may resemble certain earthquakes, especially when the number of available stations is small.
Such a misinterpretation, involving an unreported earthquake, occurred previously \citep{Selbyetal2004}.
Second, in order to detect cylindrical waves, travel time measurements are involved, which presents a 
by-no-means trivial task.
When the trajectory is unknown, it is difficult to align seismic stations, which further increases the challenge of picking the relevant arrivals.
In contrast, the linear travel-time feature associated with simultaneous global arrivals can be simply revealed by comparing seismograms side by side
in a so-called record section.
Because the signals have the same arrival time, how we align the related seismograms is less critical.
Third, due to reverberations of the CMB- and ICB-induced spherical waves,
unique free-oscillation spectra are expected.
These spectra involve a frequency-spacing between modes excited by a PBH governed by the pole-to-pole compressional-wave travel time,
and the excitation of modes not affected by earthquakes, which are restricted to Earth's crust and upper mantle.
Stacking of global seismic spectra may be required to bring out these free-oscillation characteristics.
Finally, the detector we propose applies equally to scenarios where the PBH does not collide with Earth, but passes sufficiently close to
generate detectable spherical waves. 
Of course cylindrical waves are still important, because fitting the arrival times of Mach waves provides constraints on
the PBH trajectory which cannot be obtained by considering spherical waves only.

Seismograms recorded by the global seismographic network are routinely monitored to detect landslides and glacial earthquakes \citep{Ekstrometal2003},
and it may be feasible to include detection criteria for PBHs in this monitoring process.

\section{Conclusions}\label{section:conclusions}

We study the effects of PBHs passing through or nearby Earth using numerical simulations based on a spectral-element method.
The fast motion of the PBH causes rapidly varying classical gravitational forces that are 
carefully numerically accommodated. We also satisfactorily resolve numerical issues due to the singularity in the gravitational 
near-field of the PBH.
we predict three characteristic features.
The first feature involves cylindrical waves due to a supersonic source,
generating Mach waves which are readily understood and have been studied previously  \citep{Davidetal2003}.
The second feature involves spherical waves
generated by free-slip at solid-fluid boundaries, namely the core-mantle and inner-core boundaries.
These spherical waves present a unique signature at global seismographic stations
that cannot be produced by earthquakes.
Third, free-oscillation spectra are predicted to exhibit evenly-spaced peaks, with a spacing determined by the pole-to-pole
compressional-wave travel time.
Modes not excited by earthquakes, such as $_2S_2$,
should be detectable in spectra induced by PBHs.
Probably the most likely scenario is the passage of a PBH nearby Earth.
In such a near-miss scenario, the cylindrical-wave features vanish, but the spherical waves remain.
Therefore, we may still use global seismograms and their spectra to detect a PBH,
although it would be difficulty to determine its trajectory.
We also note that spherical waves could in principle be excited by asteroids along
trajectories proximal to Earth, potentially containing new seismic information about the interior.

In this article we only consider gravitational interactions between a PBH and Earth.
Potential PBH impact phenomena, which are still debated, are not considered.
Related seismic effects may be readily taken into consideration,
as exemplified by the meteorite impact study of \cite{Mesetal2011}.
Such effects are of course irrelevant for near misses.

\bibliography{REFERENCES}

\clearpage
\appendix

\section{Convergence Tests}
\label{appendix:convergent_tests}

As discussed in Section~\ref{section:black_hole_singularity}, we introduce a critical length scale $R_\mathrm{crit}$
in order to eliminate unphysical singularities.
This length scale determines the maximum force we
allow in the numerical simulations, as well as the behavior of the source time function.
On one hand, amplitudes of simulated seismograms will be too small if
$R_\mathrm{crit}$ is too large --- the forces the PBH exerts on Earth
would be severely underestimated.
On the other hand, the smaller $R_\mathrm{crit}$, the more the effective source time function looks like a delta function,
which contains high-frequency information that cannot be resolved by our finite-size elements.
The second issue is less problematic, since high-frequency noise can be safely removed by applying low-pass or
band-pass filters in post processing.
For these reasons, $R_\mathrm{crit}$ should be chosen as small as possible, as long as
the maximum force does not exceed the limits of computer representation.
However, we will show that when the critical length scale is far less than the size of the elements,
differences in the weak form implementation vanish.

\begin{figure}[h]
\begin{minipage}{0.40\textwidth}
\centering
\includegraphics[width=0.7\textwidth,angle=0]{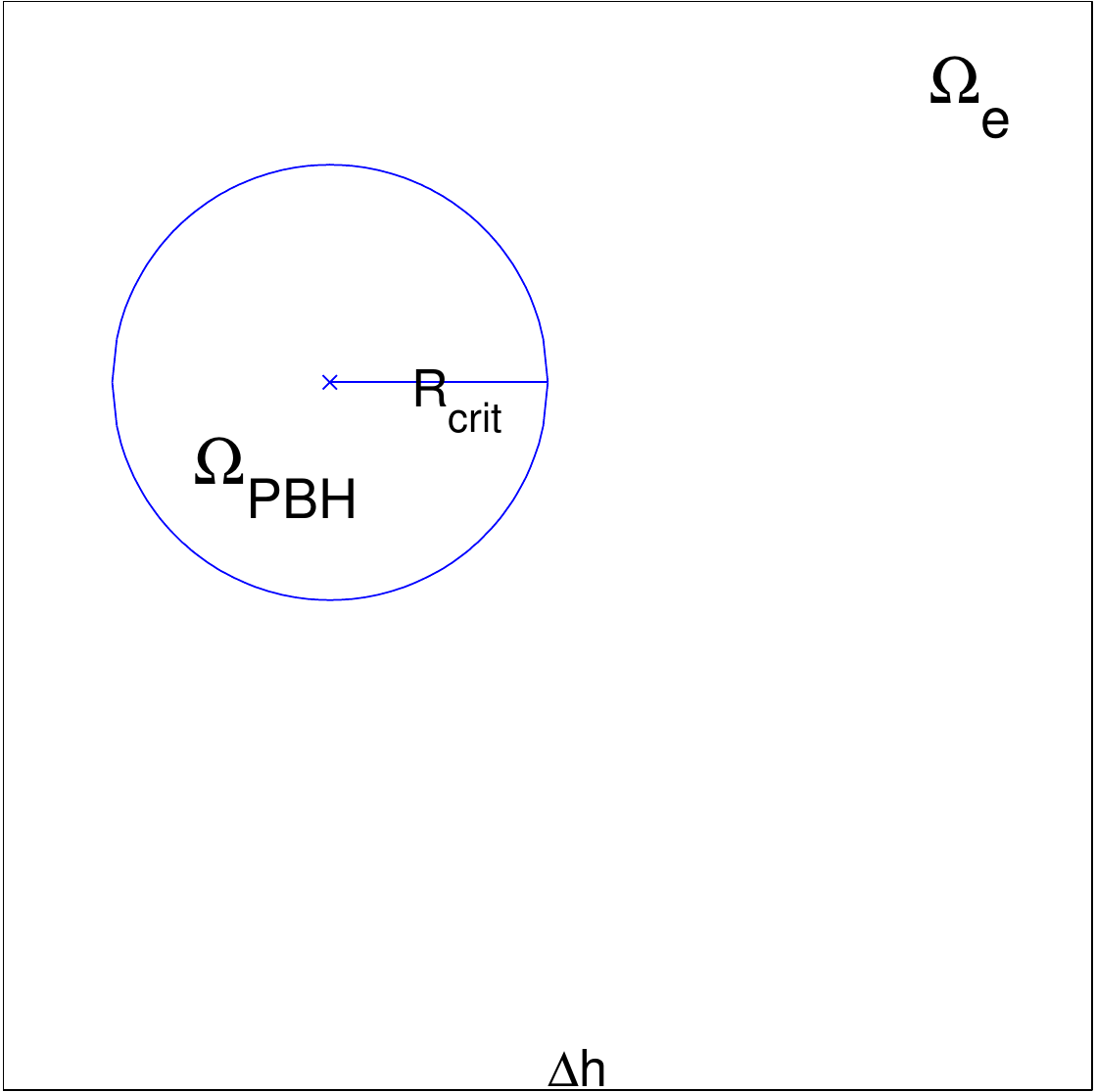} \\
\end{minipage}
\begin{minipage}{0.55\textwidth}
\caption[fig:singularity]
{
Illustration of how to avoid unphysical, singular gravitational forces from the PBH.
The big square $\Omega_e$ denotes the volume of an element and
the circle represents a tiny sphere $\Omega_\mathrm{PBH}$ around the PBH defined by the radius $R_\mathrm{crit}$.
Note that the cartoon is not drawn to scale --- the size of the sphere is exaggerated for clear viewing.
In fact, $R_\mathrm{crit}$ should be much smaller than the size of the element $\Delta h$.
The net gravitational force within $\Omega_\mathrm{PBH}$ vanishes, as long as $R_\mathrm{crit}\ll\Delta h$.
Therefore, we can avoid singularities by using the alternative force (\ref{eq:force_implementation1}).
The potential (\ref{eq:force_implementation2}) is then derived from the corresponding force.
}
\label{fig:singularity}
\end{minipage}
\end{figure}

Fig.~\ref{fig:singularity} illustrates an element which contains the PBH at an arbitrary moment.

The source term involves a volume integral of the gravitational force over element $\Omega_e$, that is
\begin{eqnarray*}
&&\int_{\Omega_e}~\bff\cdot\bfw~\rmd^3\bfx\nonumber\\
&=&\int_{\Omega_e}~\widetilde\bff\cdot\bfw~\rmd^3\bfx + \int_{\Omega_\mathrm{PBH}}\left(\bff-\widetilde\bff\right)\cdot \bfw~\rmd^3\bfx\nonumber\\
&=& \int_{\Omega_e}~\widetilde\bff\cdot\bfw~\rmd^3\bfx
+\int_{\Omega_\mathrm{PBH}}~G~M^\mathrm{PBH}~\rho(\bfx)~
\left[-\frac{\bfx-\bfx^\mathrm{PBH}(t)}{~||\bfx-\bfx^\mathrm{PBH}(t)||^3}+\frac{\bfx-\bfx^\mathrm{PBH}(t)}{R_\mathrm{crit}^3}\right]
\cdot\bfw(\bfx)~\rmd^3\bfx\nonumber \\
&\approx& \int_{\Omega_e}~\widetilde\bff\cdot\bfw~\rmd^3\bfx
+\int_{\Omega_\mathrm{PBH}}~G~M^\mathrm{PBH}~\rho\left(\bfx^\mathrm{PBH}(t)\right)~
\left[-\frac{\bfx-\bfx^\mathrm{PBH}(t)}{~||\bfx-\bfx^\mathrm{PBH}(t)||^3}+\frac{\bfx-\bfx^\mathrm{PBH}(t)}{R_\mathrm{crit}^3}\right]
\cdot\bfw\left(\bfx^\mathrm{PBH}(t)\right)~\rmd^3\bfx\nonumber \\
&=& \int_{\Omega_e}~\widetilde\bff\cdot\bfw~\rmd^3\bfx
+G~M^\mathrm{PBH}~\rho\left(\bfx^\mathrm{PBH}(t)\right)~\bfw\left(\bfx^\mathrm{PBH}(t)\right)\cdot\int_{\Omega_\mathrm{PBH}}~\left[-\frac{\bfx-\bfx^\mathrm{PBH}(t)}{~||\bfx-\bfx^\mathrm{PBH}(t)||^3}+\frac{\bfx-\bfx^\mathrm{PBH}(t)}{R_\mathrm{crit}^3}\right]~\rmd^3\bfx \quad,
\nonumber\\
\end{eqnarray*}
where we have used the fact that when $R_\mathrm{crit}\ll\Delta h$,
$\Omega_\mathrm{PBH}$ can be treated as a single point
compared to element $\Omega_e$, i.e., $\rho(\bfx)\approx\rho\left(\bfx^\mathrm{PBH}(t)\right)$ and $\bfw(\bfx)\approx\bfw\left(\bfx^\mathrm{PBH}(t)\right)$ for any $\bfx\in\Omega_\mathrm{PBH}$ at time $t$.
\newline

Define $\bfr\equiv\bfx-\bfx^\mathrm{PBH}(t)$, and hence $r\equiv|| \bfr ||=|| \bfx-\bfx^\mathrm{PBH}(t) ||$.
Then the volume integral reduces to
\begin{eqnarray}
&&\int_{\Omega_e}~\bff\cdot\bfw~\rmd^3\bfx\nonumber\\
&=& \int_{\Omega_e}~\widetilde\bff\cdot\bfw~\rmd^3\bfx
+G~M^\mathrm{PBH}~\rho\left(\bfx^\mathrm{PBH}(t)\right)~\bfw\left(\bfx^\mathrm{PBH}(t)\right)\cdot\int_{\Omega_\mathrm{PBH}}~\left(-\frac{\bfr}{r^3}+\frac{\bfr}{R_\mathrm{crit}^3}\right)~\rmd^3\bfx\nonumber\\
&=& \int_{\Omega_e}~\widetilde\bff\cdot\bfw~\rmd^3\bfx\quad,
\label{eq:effective_force}
\end{eqnarray}
because the integral over~$\Omega_\mathrm{PBH}$ is non-singular and simply vanishes.
This can be easily understood from a physical perspective: within a small spherical volume around the PBH, the net gravitational force
from the PBH is zero because gravity is isotropic in the tiny sphere.
Equation (\ref{eq:effective_force}) is important, because it provides a way to calculate
the source term involving a singular force $\bff$ using a non-singular force $\widetilde\bff$, which is easily
implemented in the spectral-element method.

\begin{figure}[h]
\centering
\includegraphics[width=0.80\textwidth,angle=0]{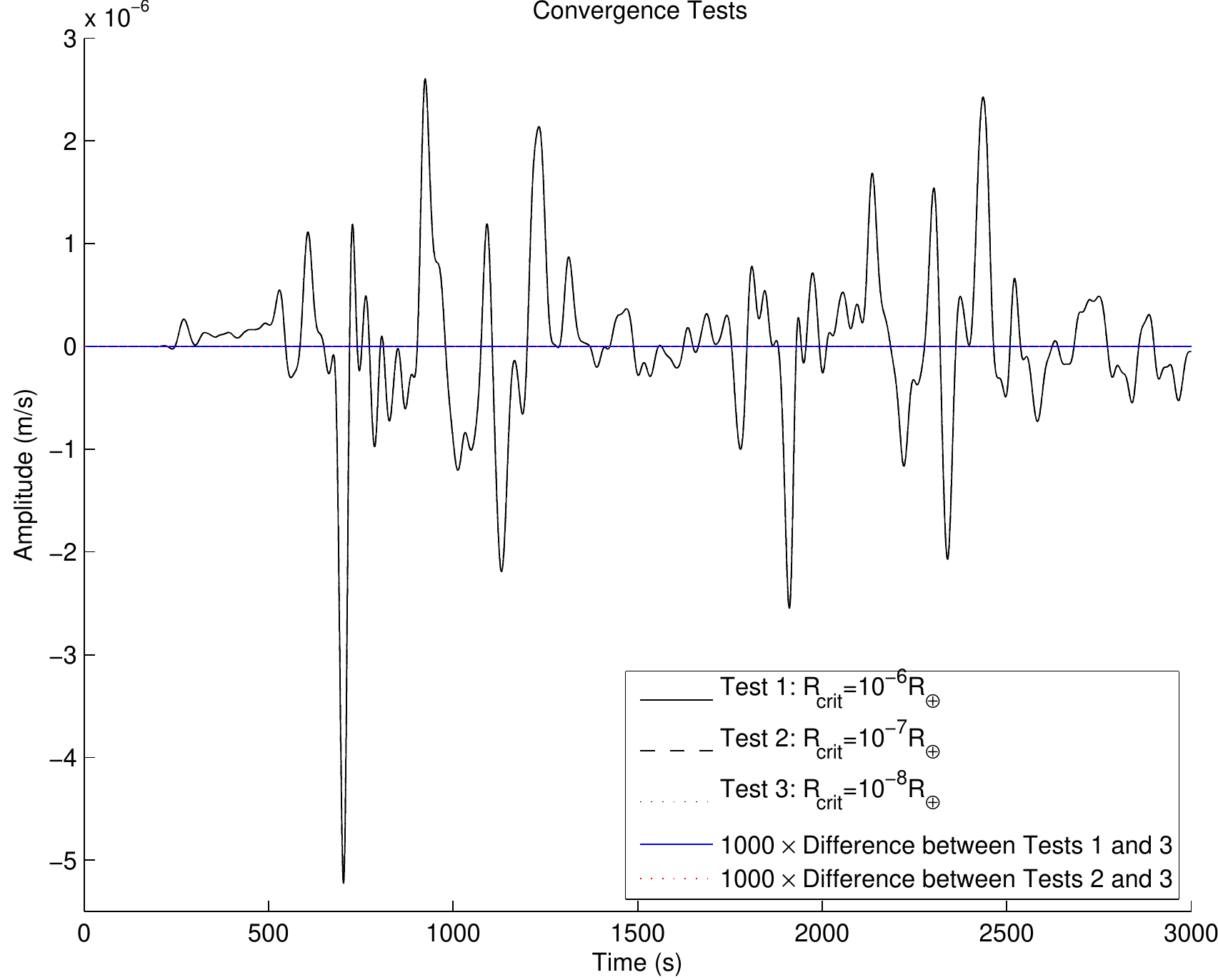} \\
\caption[fig:convergent_test]
{
Convergence tests using three different $R_\mathrm{crit}$.
The perfect match shows that the critical length scale we used for our simulations is sufficiently small.
Further decreasing $R_\mathrm{crit}$ does not change the simulation results, as expected from equation (\ref{eq:effective_force}).
}
\label{fig:convergent_test}
\end{figure}

We design an experiment to test this result.
Three complementary simulations are carried out, each with a different $R_\mathrm{crit}$: the first one with $R_\mathrm{crit}=10^{-6}R_\oplus$, the second one  with $R_\mathrm{crit}=10^{-7}R_\oplus$ and the last one  with $R_\mathrm{crit}=10^{-8}R_\oplus$, where $R_\oplus=6371~\mathrm{km}$ is the radius of Earth.
One seismogram from each of the three simulations is shown in Fig.~\ref{fig:convergent_test}.
The three simulations give the same results, thereby demonstrating
 that the critical length scale $R_\mathrm{crit}=10^{-6}R_\oplus$ used in our simulations is sufficiently small for our
theoretical derivation to be valid.
Using $\widetilde\bff$ instead of $\bff$ does not only removes singularities, but also gives reliable seismograms.

\begin{figure}[h]
\centering
\includegraphics[width=0.70\textwidth,angle=0]{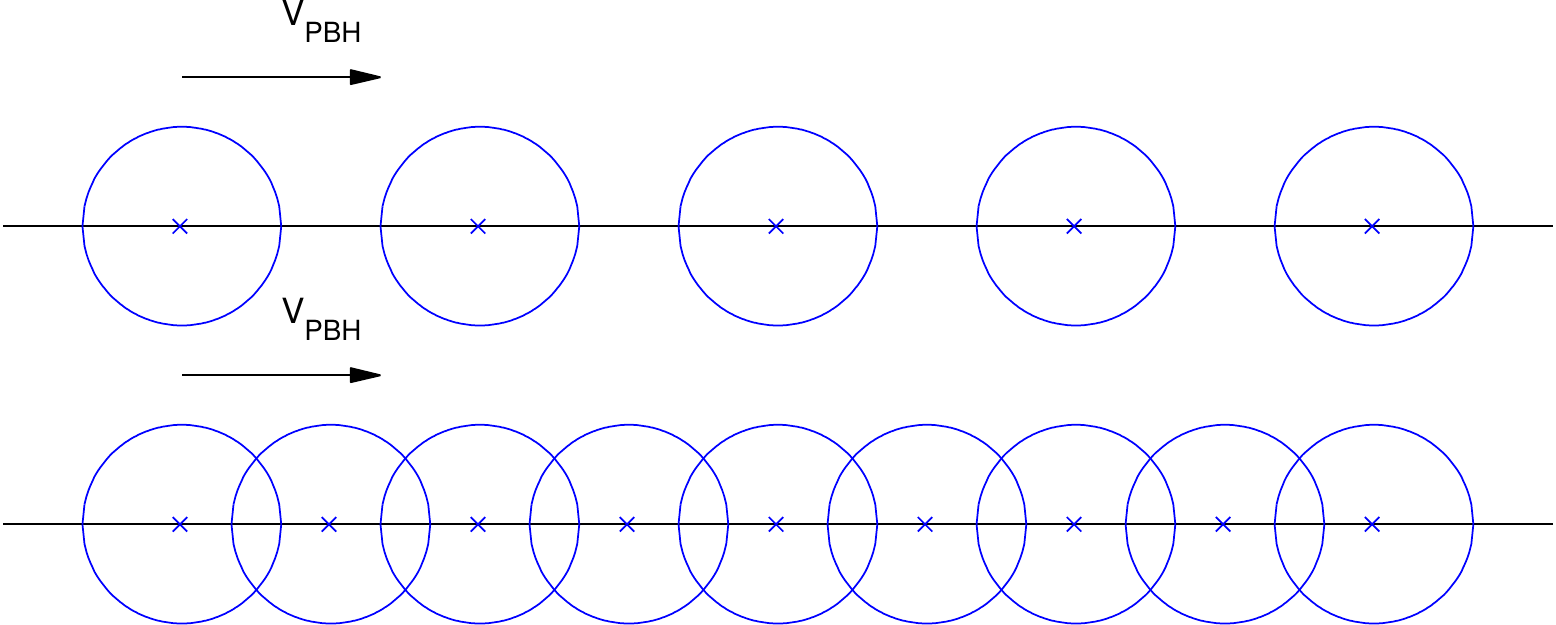}\\
\caption[fig:resolve_trajectory]
{
Illustration of the significance of the choice of time step $\Delta t$. Two situations are shown, where the only difference is that
$\Delta t$ for the top row is twice that of the bottom row. The straight lines represent the trajectories of the PBH, whereas the
crosses denote the PBH location at incremental time steps. The blue circles highlight the effective region of the PBH,
i.e., the gravitational forces from the PBH outside these regions may be neglected because of the $1/r^2$ fall-off of the gravitational field.
Note that the radius of the effective region is not identical to the critical length scale $R_\mathrm{crit}$.
When $\Delta t$ is chosen too large (top), there are points along the trajectory that are not sampled by the time steps.
Therefore, the effect of the PBH is greatly underestimated. When $\Delta t$ decreases (bottom), we cover the
entire trajectory, which fully accounts for the gravitational effects of the PBH.
}
\label{fig:resolve_trajectory}
\end{figure}

Another issue involves capturing the widely differing time scales associated with the motion of the PBH and the periods of  the induced seismic waves.
Fig.~\ref{fig:resolve_trajectory} explains why this is important.

\clearpage
\section{More Examples}
\label{appendix:more_examples}

\begin{figure}[h]
\centering
\includegraphics[width=0.98\textwidth,angle=0]{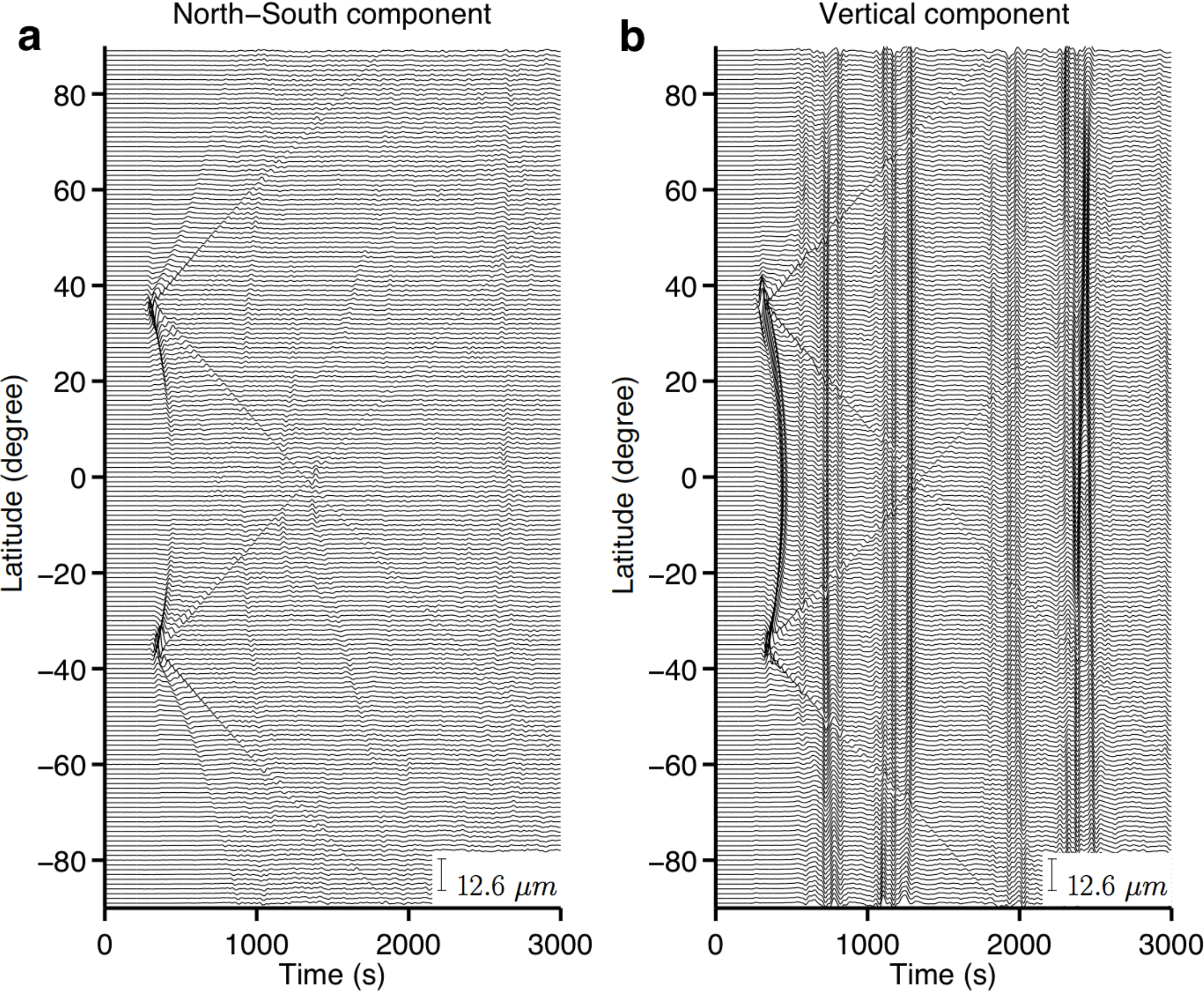} \\
\caption[fig:trajectory3]
{
Seismograms from a simulation in which $d=0.8~R_\oplus$ and $v=200$~km~s$^{-1}$.
Direct arrivals and Rayleigh surfaces waves are observed on both horizontal and vertical components,
whereas phases that share the same travel time at all stations (straight vertical features in the figure on the right) only emerge on the vertical component.
The focusing points of the surface waves are determined by the entrance and exit points of the PBH,
while the different curvatures of first arrivals at high and low latitudes are an expected feature of Mach waves.
}
\label{fig:trajectory3_200}
\end{figure}
\clearpage

\begin{figure}
\centering
\includegraphics[width=0.98\textwidth,angle=0]{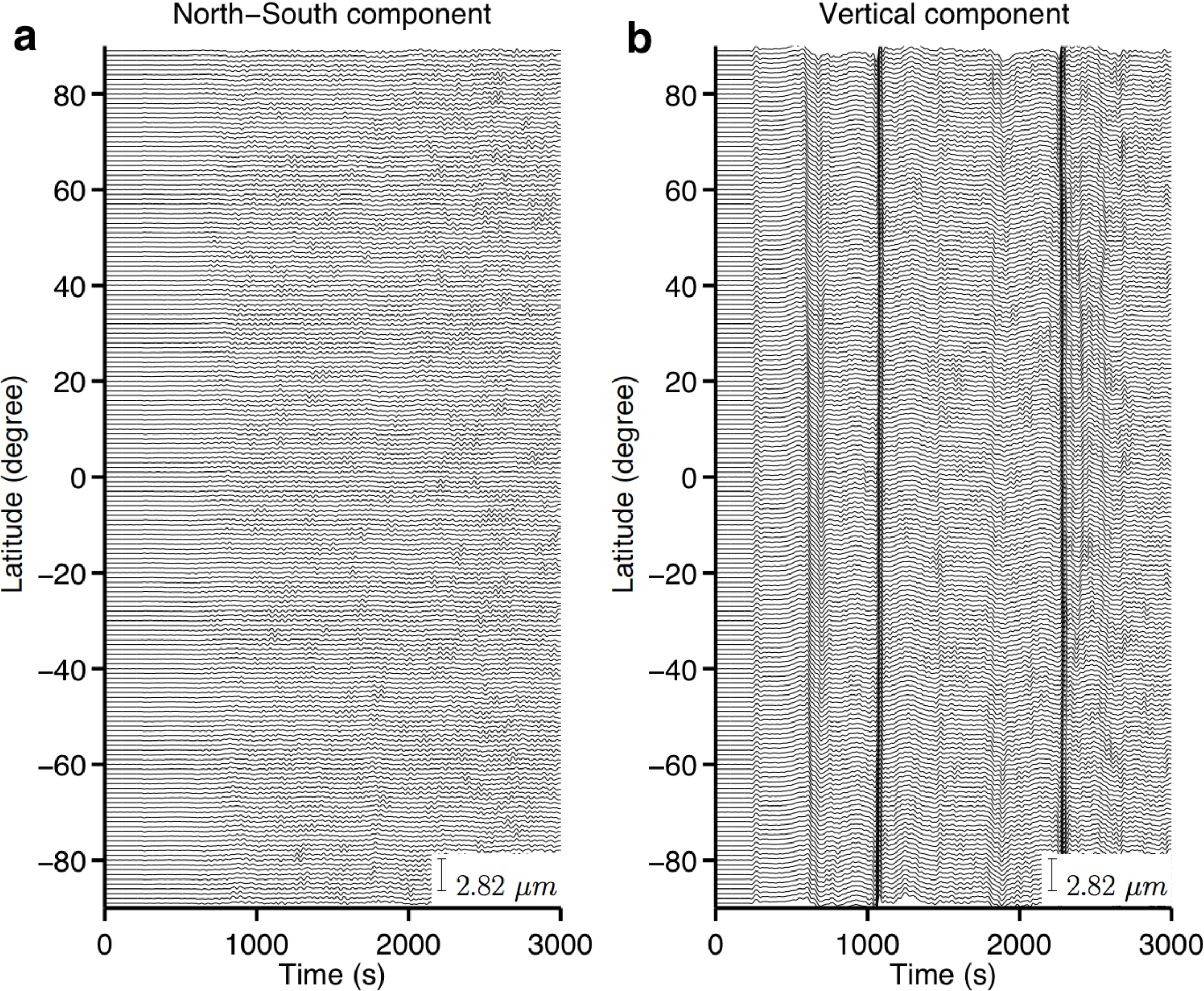} \\
\caption[fig:trajectory5]
{
Seismograms from a simulation in which $d=1.6~R_\oplus$ and $v=200$ km s$^{-1}$.
The key observation for this scenario is that no coherent signals are observed on the North-South component,
because the PBH does not transit Earth.
When there is no direct contact between the PBH and Earth,
the dominant mechanism changes from a supersonic source to differential gravitational forces at different locations in
Earth's interior, a result which is more difficult to imagine and interpret.
Nevertheless, not only does this result validate our explanations for the cylindrical- and spherical-wave features,
it also provides a new way to detect a PBH, even when it passes nearby Earth.
}
\label{fig:trajectory5_200}
\end{figure}
\clearpage

\begin{figure}
\centering
\includegraphics[width=0.98\textwidth,angle=0]{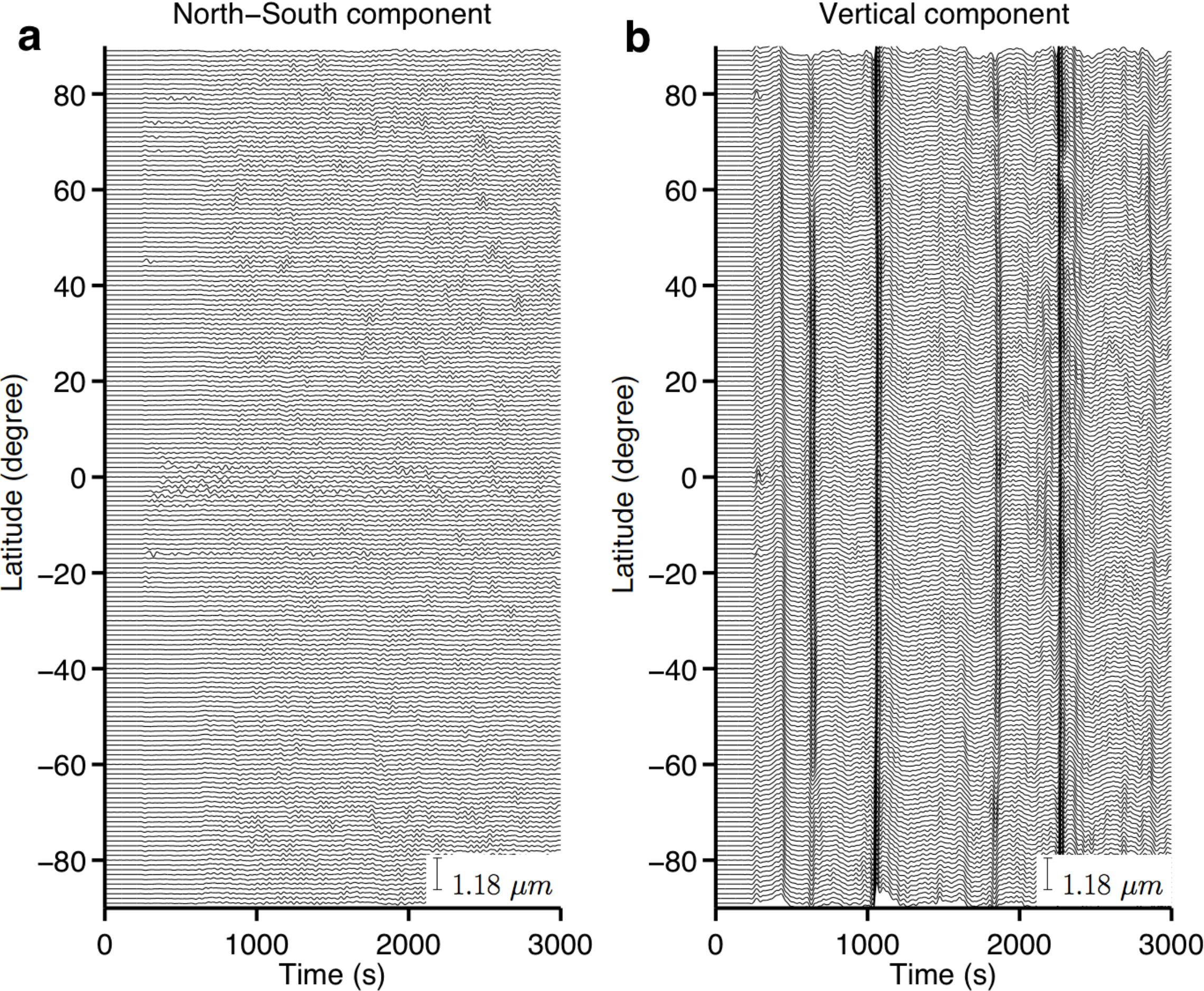} \\
\caption[fig:3D]
{
Simulation result for 3-D Earth model S40RTS with $d=1.6~R_\oplus$ and $v=400$~km~s$^{-1}$.
The characteristic signature from the spherical waves remains the same.
}
\label{fig:3D_earth}
\end{figure}

\clearpage
\renewcommand{\thefigure}{\arabic{figure}}
\setcounter{figure}{0}

\end{document}